\documentstyle [12pt] {article}
\def\lsim{\lower.5ex\hbox{$\; \buildrel < \over \sim \;$}}
\def\gsim{\lower.5ex\hbox{$\; \buildrel > \over \sim \;$}} 
\def \simeq{\lower.3ex\hbox{$\; \buildrel \sim \over - \;$}}
\title{\sc Spectral Properties of Accretion Disks Around
Galactic and Extragalactic Black Holes}
\author{Sandip K. Chakrabarti$^1$\\
{\small Code 665, Goddard Space Flight Center, Greenbelt, MD, 20771}\\
and\\
Lev G. Titarchuk\\
{\small George Mason University, 4400 University Drive, Fairfax, 22030}\\
{\small and}\\
{\small Code 668, Goddard Space Flight Center, Greenbelt, MD, 20771}\\
}
\begin{document}
\noindent Received 1995 May 16;  Accepted 1995 June 27; Publication date
1995, Dec. 20, v. 455, p 623, Astrophysical Journal \\

\noindent Following erratum submitted 29th January:

While reading the description of Figure 2, as well as the 
caption of the same Figure, it is to be noted that 
the solid, dashed, short dashed and dotted lines were drawn for ${\dot m}_d= 0.001$, $0.01$, $0.1$ and $1.0$ respectively. The sequence in
inadvertantly reversed in the text.

\maketitle
\pagenumbering{arabic}
\begin{abstract}
\baselineskip 22pt
We study the spectral properties of a very general class of accretion disks
which can be decomposed into 
three distinct components apart from a shock at $r=r_s$: 
(1) An optically thick Keplerian disk on the equatorial plane ($r>r_s$), 
(2) A sub-Keplerian optically thin halo above and below this Keplerian
disk $r>r_s$ and
(3) A hot, optically slim, $\tau\sim 1$ postshock region $r<r_s\sim 5-10 r_g$
where $r_g$ is the Schwarzschild radius. The postshock region
intercepts soft photons from the Keplerian component and reradiates
them as hard X-rays and $\gamma$ rays after Comptonization. 
We solve two-temperature 
equations in the postshock region with Coulomb energy exchange between
protons and electrons, and incorporating radiative processes such as
bremsstrahlung and Comptonization. We also present the exact prescription
to compute the reflection of the hard X-rays from the cool disk.
We produce radiated spectra from both the disk components as 
functions of the accretion rates and compare 
them with the spectra of galactic and extragalactic black hole candidates.
We find that the transition from hard state to soft state is
smoothly initiated by a single parameter, namely, 
the mass accretion rate of the disk. In the soft state, when the postshock
region is very optically thick and cooled down, bulk 
motion of the converging flow determines the spectral index
to be about $1.5$ in agreement with observations.

\end{abstract}

{\it Subject headings}:\ {\rm accretion, accretion disks -- black hole
physics -- radiation mechanisms:nonthermal -- shock waves --stars:neutron
}\\

\noindent $^1$ NRC Senior Research Associate; Address after October, 1995:
 Tata Institute of Fundamental Research, Homi Bhabha Road, Colaba, 
Bombay, 400005, INDIA\\

\noindent {e-mails: chakraba@tifrvax.tifr.res.in; 
titarchuk@lheavx.gsfc.nasa.gov}

\baselineskip 22pt
\newpage

\noindent {\Large 1. INTRODUCTION}

In the 1970s, the standard accretion disk models were constructed
(Shakura \& Sunyaev 1973; Novikov \& Thorne 1973)
to largely explain observations of binary systems. Matter supply
from the companion is assumed to be through the Roche lobe and the 
angular momentum distribution is assumed to be Keplerian throughout the
disk. Here, viscosity drives the inflow by removing angular momentum outward
and keeping the entire disk Keplerian in the process. In the early 1980s,
these disks became favorite candidates for the explanation of the
``big blue bump'' seen in the UV region of the active galaxies
(Malkan 1982; Sun \& Malkan 1989 and references therein) with some success.

With the advent of space based observations, the X-ray and $\gamma$-ray
spectra of these objects show that there could be even bigger
bumps in the X-ray regions of the spectra. Galactic
black hole candidates such as GS1124-68, show two distinct spectral
components, namely, soft and hard, which apparently vary quite 
independently (Tanaka 1989, hereafter TAN; Ebisawa et al., 1993, 1994 hereafter
E93 and E94, respectively; Inoue 1992, hereafter I92). 
It was becoming clear that the standard accretion 
disks, which are in general cooler, cannot explain the production
of X-rays and $\gamma$-rays. These models also cannot explain 
the almost zero time-lag correlated variabilities 
observed in, e.g., NGC4151, NGC5548 (Clavel et al. 1990;
Peterson et al. 1991; Perola et al. 1986).
Explanations of the hard components seen in the
spectra of neutron star candidates and galactic and extragalactic black hole
candidates required extra components, such as plasma clouds, hot coronae etc. 
which are not self-consistently constructed. 

That the disks need not be of ``standard'' type was sensed by theoreticians
since early 1980s (Liang \& Thomson, 1980; Paczy\'nski \&
Bisnovatyi-Kogan 1981; Muchotrzeb \& Paczy\'nski 1982; Paczy\'nski \& Wiita
1980; Abramowicz et al. 1988, Chakrabarti 1989, hereafter C89; Chakrabarti
1990ab, hereafter C90ab, Chakrabarti \& Molteni 1995, hereafter CM95).
Particularly, tt was realized that the disks are not Keplerian
everywhere, especially at the inner boundary.
(For a general discussion see, Chakrabarti 1995a, hereafter C95a).
First global and fully self-consistent solution of viscous,
transonic flows was obtained by C90ab who found that
a Keplerian flow at the outer boundary can become sub-Keplerian
at the inner edge (with or without shock waves). Or, contrarily,
an initially sub-Keplerian flow can become almost Keplerian when
viscosity is high enough. In a binary system, which includes a black hole
or a neutron star, matter could be accreted from the winds of the
companion at the same time as it is through the Roche lobe.
In low mass X-ray binaries (LMXB), where companion winds may be scarce
(except probably during outburst phase)
the sub-Keplerian flow could come from the Keplerian disk itself
as the advection becomes important closer to the black hole. In an active
galaxy, the incoming matter is either a mixture of Keplerian and sub-Keplerian
flows, or, entirely sub-Keplerian. This is because matter may be supplied
from constantly colliding winds from a large number of stars $\sim$ pc away
and thus most of the angular momentum of the infalling matter may be lost. 
The above solutions, obtained in the context of isothermal flows, have been
generalized since then for any heating and cooling (C95a; Chakrabarti 1995b,
hereafter C95b) but no new topologies emerged.

After the matter starts with sub-Keplerian or Keplerian angular momentum $l
\leq l_{Kep} (r_{out})$, its subsequent behavior depends strongly 
upon the accretion rate and the viscosity in the flow (C90ab; CM95, 
C95ab, Appendix). A unified description of what the flow might do (e.g., Fig. 4
of C89) suggests that, for a given entropy of the flow, 
if the accretion rate is small enough, the flow passes
through the outer sonic point (just as in a Bondi flow) and remains supersonic
before falling onto a black hole. If the accretion rate is high, 
matter would pass through the inner sonic point. In the intermediate
case, matter is virtually stopped at a distance $r_s$, where $l \simeq
l_{Kep} (r_s)$ due to the centrifugal barrier, and a shock is formed. 
The postshock flow becomes highly subsonic momentarily, but picks up 
its velocity as it approaches the black hole eventually entering the horizon 
supersonically. The postshock region behaves similar to the thick
accretion disks discussed in the literature (e.g., Paczy\'nski \& Wiita
 1980)
but is more self-consistent as advection is explicitly added (C89, Molteni,
Lanzafame, \& Chakrabarti 1994, hereafter MLC94). In these solutions, 
the entropy of the flow passing through the inner sonic point is higher 
compared with the entropy of the flow passing through the outer sonic
point. The combined effects of the
shock and the viscosity generate the right amount of entropy to allow
the flow to pass through the inner sonic point. The energy of the flow
is either constant or very nearly so, because the energy and the entropy
are advected with the flow toward the black hole. The
advection is clearly important, since the radial velocity could be
very high close to the black hole. When accretion rate is low
enough, and shock conditions are not satisfied, a Keplerian
disk would become sub-Keplerian close to the black hole
at the same time becoming gas pressure dominated (Rees et al., 1982).
Recently, importance of advection is stressed (`newly discovered
advection dominated flows') by a simple self-similar analysis by
Narayan \& Yi (1994) who analyzes one of the branches of 
the solution topologies found by C90a,b.

The most general and complete global solution to-date
of the viscous sub-Keplerian flow with advection (which becomes
Keplerian at  a larger distance),
shows a very interesting behavior (C90a,b, CM95, C95b). 
When the viscosity is low [typically, when the viscosity
parameter (Shakura \& Sunyaev 1973)
is less than a critical value, $\alpha_v <\alpha_{vc}$],
the shock can still form (depending on accretion rate, see C89) as in an
inviscid flow, but as the viscosity in increased ($\alpha_v>\alpha_{vc}$), 
the stable shock becomes
weaker until it  finally disappears. Numerical simulation
provides insight into what actually happens (CM95). In the presence of
viscosity, the rate of transport of angular momentum is
faster in the postshock subsonic flow as compared with the 
preshock supersonic flow. As a result, angular momentum starts `piling up'
in the postshock flow and pushes the shock away to a larger distance.
If one employs the Shakura-Sunyaev (1973) prescription,
and the viscosity is small enough ($\alpha_v \lsim \alpha_{vc} \sim 0.01$
for isothermal disk, C90a), the transport
rates could be matched and the shock settles down farther out. If the viscosity
is higher, the transport rate is fast enough to push the shock
to extinction and the flow becomes subsonic and Keplerian except
close to the inner edge of the disk, where it passes through the
inner sonic point. In the case of
more general cooling and heating laws, $\alpha_{vc}$ could be anywhere
between $0$ to $1$ depending upon the relative importance of
cooling and heating (C95a,b). The predictions of shock locations and other 
properties from our vertically averaged model were found to agree with three
dimensional simulations (MLC94). We, therefore, believe that these solutions
are generally applicable even for quasi-spherical flow. This has also been
noticed by recent self-similar study of Narayan \& Yi (1994).

Numerical simulations (MLC94, see also Hawley et al. 1984, 1985; Ryu et al. 
1995) also show that a strong wind forms in the postshock flow. Analytical
works (C89) show that winds form when the specific energy of
matter is positive and the specific entropy is high.
The wind is found to become supersonic within a finite distance (MLC94). 
In the unified description (Fig. 4 of C89), the region of parameter space 
which produce winds (with and without shock waves) is discussed as well.

These, exact and
global solutions (C90a,b, C95a,b) with higher viscosity,  of the original 
model of transonic viscous disks of Paczy\'nski and collaborators
represent the most complete, Keplerian and quasi-Keplerian
accretion disks in the literature. It has been suggested 
(Chakrabarti, 1993, hereafter C93, Chakrabarti, 1994, hereafter C94, C95a,b)
that they go over to the so-called ``thin'', ``thick'', ``slim'' and 
``newly discovered advection dominated'' and ``cooling dominated'' disks
when appropriate approximations are made. Moreover, the solution of 
transonic viscous disk (C90a,b, CM95, C95a,b) also includes 
stationary shock waves for lower viscosity and appropriate
accretion rate limit (C89). This `grand unification' of disk models
was possible, because the most general equations with
all the solution branches including shocks are used (C93, C94, C95a,b) 
without making any further restrictive assumptions
such as self-similarity (Narayan \& Yi, 1994) or Keplerian
distribution (Chen et al., 1995). Based on these solutions, one can construct
the most general accretion disk close to a black hole.
Assume for the moment that the flow is sub-Keplerian at the outer boundary
and assume that the viscosity is higher on the equatorial plane
and gradually drops with vertical height. The higher viscosity in the 
equatorial region re-distributes the sub-Keplerian angular momentum
and makes a standard, optically thick,
Keplerian disk on the equatorial plane (CM95). However, the smaller 
viscosity at higher latitudes would produce an optically thin halo and this gas
will pass through a shock close to the black hole, forming a very hot
corona. In cases when an admixture of Keplerian and sub-Keplerian
matter is supplied at the outer boundary (e.g., when both 
Roche lobe accretion and wind accretion are present) the segregation of this
flow into two components is trivially achievable. 
In case where only Keplerian matter is supplied at the outer boundary,
the disk can  still be segregated in the same manner closer
to the black hole depending on the viscosity variation and its
vertical distribution which determines the region from which the flow
starts deviating from Keplerian (C90a,b, Appendix).
The solution with the shock (if available)
is always chosen in place of the shock-free solution, as the former 
has a higher entropy (C89, MLC94).

Fig. 1 shows a cartoon diagram of the generalized accretion disk in 
presence of Keplerian and sub-Keplerian matter close to a black hole 
(also see, C94). An optically thick Keplerian disk is flanked by an
optically thin, sub-Keplerian halo which 
passes through a standing shock close to the
centrifugal barrier. The postshock region forms a hot
ion pressure supported torus (e.g. Rees et al., 1982; MLC94)
of moderate optical depth $\tau \sim 1$
as the radial kinetic energy of the infall is 
thermalized: $\rho v_r^2(r_s) \sim P \sim n_H k T_p$.
For matter with marginally bound angular momentum,
the shock forms around $r_s \sim 10r_g$ ($r_g\!=\!{2GM/c^2}$,
the radius of the black hole) in Schwarzschild geometry.
The flow remains dissipative up to the marginally stable radius 
$r_{ms}\!=\!3r_g$, beyond which the flow rapidly passes through the sonic
point. For a flow around a Kerr black hole of angular momentum
parameter $a\!=\!0.99$, these length scales are roughly half as long.
(Henceforth, we choose units of distance and velocity to be 
$r_g$ and $c$ respectively. We continue to use $r$ to denote
radial distances with this unit.) For higher angular momenta, shocks form 
farther away (C89). The postshock flow becomes geometrically thick,
intercepting a few ($\lsim 5$) percent of the 
soft photons from the optically thick, Keplerian
component of the preshock accretion disk. The postshock region with a 
Thomson optical depth of $\tau_h \sim \int_{x_{ms}}^{x_s} 0.4 \rho dx $ 
($\sim 1\!-\!3 $ for ${\dot M} \sim {\dot M}_{Edd}$ rates)
heats up these soft photons through inverse Compton scattering 
(Sunyaev \& Titarchuk 1980, hereafter ST80; Titarchuk 1994, hereafter T94; 
Titarchuk \& Lyubarskij 1995; hereafter TL95 respectively.)
and the photons are re-emitted which we claim to be
responsible for the hard radiations from the accretion disks
around galactic and extragalactic black 
holes. Some of these hard photons are intercepted by the Keplerian
disk, a part of which is absorbed, thermalized, and reradiated by the disk
as soft radiations, while the rest is reflected. We
compute, completely self-consistently, the fraction of incoming
hard radiation that each annulus absorbs or reflects away. 
Based upon our analysis, we do not find the component of hard
radiation reflected from the disk to contribute significantly to the
observed spectra. Secondly,
we find the Keplerian disk component is not suitable for the
site of the iron line emissions. Rather, we believe that the outflowing
winds from the postshock flow (MLC94) or the evaporated
preshock disk, could be the site of these lines at various stages 
of ionization (Chakrabarti et al., 1995, hereafter 
CTKE95). We shall touch upon this aspect of the problem in the final 
section. When the net accretion rate is high enough, cooler photons 
trapped in our postshock region are Comptonized by the `converging flow' 
(e.g. Blandford \& Payne, 1981) for
$r \leq r_{convg} \sim 3 ({\dot M}_d + {\dot M}_{h}) / {\dot M}_{Edd}
r_g$ where, ${\dot M}_d$ and ${\dot M}_h$ are the disk and halo
rates respectively, and ${\dot M}_{Edd}$ denotes the Eddington
accretion rate. The resulting hard spectrum is very weak with a
spectral slope of $\alpha \sim 1.5$  ($F_\nu \propto \nu^{-\alpha}$)
and is characteristics of the spherical converging optically thick
flow (Titarchuk, Mastichiadis \& Kylafis 1995, hereafter TMK96).

The advantage of this model is that the soft and the hard radiations
are formed self-consistently from the same accretion disk without invoking
any ad hoc components such as the plasma cloud, hot corona, etc., whose
origins have never been made clear. Secondly, our results are
sensitive to only one parameter, namely, the disk accretion rate ${\dot m}_d$,
although some observations suggest a variation of the halo accretion
rate ${\dot m}_h$ cannot be ruled out. 
In the next Section, we present the equations for the hydrodynamics
and radiative processes which we use to describe our model. In \S  3,
we present the results and physical interpretations of our solution.
In \S  4, we present general comparisons with observations. In \S  5, 
we make concluding remarks.

\noindent{\Large 2. THE MODEL DESCRIPTION AND THE BASIC EQUATIONS}

\noindent {\large 2.1 Description of the Model}

As mentioned in the introduction, our model consists of two major disk
components:

\noindent (1) The Standard Disk Component: A standard, optically thick
disk could be produced either from the Keplerian or the
sub-Keplerian matter at the outer boundary, if the viscosity
and cooling are appropriate (C90a,b, CM95, C95a,b, Appendix). 
While constructing the equatorial Keplerian disk, 
we therefore incorporate the usual assumptions regarding the flux
emitted from the disk. We assume that before the shock
is formed at $r\!=\!r_s$, there is a thin, standard, Keplerian 
disk on the equatorial plane which, in the absence of the shock,
would emit a flux of (e.g, Shapiro \& Teukolsky 1984)
$$
F_{SS}= 7.6 \times 10^{26} r^{-3} {\cal I} (\frac{M}{M_\odot})^{2} 
\frac{\dot M}{1.4\times 10^{17}} {\rm erg\ cm^{-2}\ s^{-1}}.
\eqno(1)
$$
Here ${\cal I}\!=\! 1-(3/r)^{1/2}$. In the presence 
of the shock at $r_s$, this component continues till $r_s$. The soft
radiation luminosity  $L_{S}$ is then the integral
of the above flux from $r_s$ outwards plus the intercepted radiation 
coming from the postshock region:
$$
L_{S}= L_{SS}+L_{H} f_{sd} (1-{\cal A}_\nu) ,
\eqno(2)
$$
where, ${\cal A}_\nu$ is the albedo of the disk, $L_{H} \sim L_{SS} 
f_{ds} E_{Comp}$ is the hard
luminosity from the postshock region, $f_{sd}$ is the fraction
(determined by the geometry) of $L_{H}$ that is intercepted by the disk,
and $E_{Comp}$ is the factor by which the soft photons incident on the
postshock region (fraction $f_{ds}$ of $L_{SS}$) is enhanced due to
Comptonization. The second term therefore represents the luminosity
of the soft photons resulting from reprocessing of the
hard radiation intercepted by the Keplerian disk.

\noindent (2) Sub-Keplerian Halo Component:
We consider in detail the properties of the sub-Keplerian 
halo component and the effects of the halo on the emission
properties of the Keplerian
component. We assume the halo to be quasi-spherical and axisymmetric.
For the purpose of obtaining the height of the shock, we assumed
the flow to be in vertical equilibrium in the postshock region.
The height of the shock enables us to
compute the fraction $f_{sd}$ of disk soft photons intercepted by the shock
as well as the fraction $f_{ds}$ of hard radiations from the 
shock intercepted by the disk. We assume the polytropic index
to be $\gamma\!=\!5/3$ in the halo. A $\gamma\!=\!5/3$ spherical adiabatic flow
cannot have a shock (C90b), but if the flow is thin,
it can have two sonic points and therefore a standing shock. 
Even when a flow has only the inner sonic point,
a shock can form if it is already supersonic at the source.
An effective polytropic index is likely to be $4/3 >\gamma > 5/3$,
but we do not expect any change of our general conclusions.
Since we are interested in accreting solutions, the angular
momentum in the halo component should be around the marginally bound
and marginally stable values. In this low angular momentum regime,
in the Schwarzschild geometry, the shock may be located around 
$r_s\!=\!10-30$ or so (C89). The inner edge of the
postshock flow is extended till only the inner sonic point $r_i$, since 
for $r <r_i$, advection is so strong that matter may be considered
radiatively inefficient. Since the inner sonic point is close to the
marginally stable orbit $r_{ms}\!=\!3$ anyway, we integrate the equations 
only from $r_s$ to $r_{ms}$. We also considered the solutions
around a Kerr black hole, where $r_s \sim 5$ and $r_{ms} \sim 1.5$
will be employed. These are reasonable values for $a\sim 0.99$.
The results are similar, though vary only in details. Flow
within the inner sonic point constitutes the so-called convergent
inflow provided the accretion rate is high enough which produces
the characteristic weaker hard component as will be discussed later.

\noindent {\large 2.2 Hydrodynamics}

We solve two temperature equations for the sub-Keplerian halo component
and the postshock region.
We ignore the vertical structure of the halo and use the vertically 
integrated equations. Comparison of analytical (C89) and numerical
simulations (MLC94) show that such assumptions are justified. The radial 
momentum equation is given by,
$$
v_r \frac{dv_r}{dr} + \frac{1}{\rho}\frac{dP}{dr}-\frac{l^2}{r^3}+ \phi(r)=0,
\eqno(3a)
$$
where $P$ and $\rho$ are vertically integrated pressure and density,
respectively, $l$ is the specific angular momentum and $\phi(r)=
-1/2 (r-1)^{-1}$ denotes the Paczy\'nski-Wiita (Paczy\'nsi \& Wiita, 1980)
potential which mimics the geometry around a black hole quite well. 
The variation of the azimuthal angular momentum in presence of viscosity 
is given by
$$
v_r\frac{dl}{dr}-\frac{1}{ h r \rho}\frac{d}{dr}(r^2 W_{r\phi})=0.
\eqno(3b)
$$
We assume the viscosity in the halo to be small enough ($\alpha_v 
<\alpha_{vc}$) so that the centrifugal barrier can form a shock
in the halo (C90a,b, CM95, C95a,b). However, we assume the viscosity 
on the equatorial plane to be high enough so that the
angular momentum there is Keplerian. This would produce a generic
composite disk with optically thick (Keplerian disk), thin
(sub-Keplerian halo) and slim ($\tau\sim 1$, postshock region) components.
Whether all the three components will be present will depend on viscosity
and cooling processes.
Detailed classification of global solutions will be discussed elsewhere
(C95b).

In the region, $r<r_s$, the Keplerian and sub-Keplerian components can
behave identically for three reasons: (1) the angular momenta of the
sub-Keplerian and Keplerian components are roughly the
same at the centrifugal barrier, i.e., at $r \lsim r_s$;
(2) $v_r \approx 0$ in the immediately postshock region of the halo
whereas $dv_r/dr \approx 0$ in the disk component (velocity shoots up 
from a minimum, C89, C95b, Case A in Fig. 10 of Appendix A) --- in 
either case the advection term $v_r dv_r/dr$, in the immediate 
vicinity of the shock is negligible and both the flows accrete downstream 
as a single component, and, most importantly,
(3) since the postshock region is much hotter (more than 2 to 3 
orders of magnitude, see, \S 3) than the Keplerian component,
it evaporates the disk underneath and together they
are expected to behave as a single component (Fig. 1)
provided the mixing of the disk and halo are efficient. In the
present paper we assume that the turbulent mixing time is slower than
the infall time scale. In a neutron star accretion the mixing is complete.
This case will be discussed elsewhere.

The optically thin preshock halo does not radiate efficiently
and therefore advects the energy and entropy along with the flow exactly
as in our previous model of transonic disks (C89; C90a,b; C95b).
In the postshock region, the disk will be hotter.
If the flow is in hydrostatic equilibrium in the vertical
direction, Mach numbers at the shock are related by (C89, C90b):
$$
\frac{[M_+(3\gamma -1)+\frac{2}{M_+}]^2}{2+(\gamma - 1)M_+^2}
=\frac{[M_-(3\gamma -1)+\frac{2}{M_-}]^2}{2+(\gamma - 1)M_-^2}.
\eqno{(4a)}
$$
(The subscripts ``-'' and ``+'' denote the preshock and postshock
quantities, respectively.)
Assuming a strong shock, it is easy to show that the post
shock Mach number is $M_+ \sim 0.35$
which gives the postshock proton temperature as:
$$
T_p \sim 1.56 \times 10^{12} r_s^{-1} \ K.
\eqno(4b)
$$
However, if the flow were not assumed to be in vertical equilibrium,
but remained thin, a different Mach number relation is to be used (C90b):
$$
\frac{(M_+ \gamma + \frac{1}{M_+})^2}{2+(\gamma-1)M_+^2}=
\frac{(M_- \gamma + \frac{1}{M_-})^2}{2+(\gamma-1)M_-^2} .
\eqno{(5a)}
$$
In this case, the postshock Mach number for a strong shock is
$M_+=0.45$, and the proton temperature in the postshock region is,
$$
T_p \sim 0.96 \times 10^{12} {r_s}^{-1} \ K
\eqno{(5b)}
$$
For a spherical flow with $\gamma\sim 5/3$, shocks can form if
the accreted sub-Keplerian matter is already supersonic (this is true
for models in vertical equilibrium if $\gamma \gsim 1.5$, C95a)
but for a thinner flow even with $\gamma=5/3$ there are two saddle
type sonic points and shocks could form. Subsequent behavior of the protons
and electrons will depend on the heating and cooling processes. As the
matter accretes, it is compressed due to geometrical effects and becomes
hotter. At the same time, the electrons lose energy
due to bremsstrahlung and Comptonization of the soft photons from the disk.
They also gain energy from the protons due to the Coulomb interactions, and
cool the protons in the process.  The energy equation which the protons and
electrons obey in the postshock region is:
$$
\frac{\partial}{\partial r} (\epsilon + \frac{P}{\rho}) + (\Gamma-\Lambda)=0 ,
\eqno(6)
$$
where $\epsilon$ is given by,
$$
\epsilon=\frac{1}{2}v_r^2 + \frac{P}{\rho(\gamma-1)}-\frac{1}{2(r-1)}
+\frac{l^2}{2r^2},
\eqno(7)
$$
and $\Gamma$ and $\Lambda$ are the heating and cooling terms respectively.

In the equation which governs the proton energy, 
we consider $\gamma=5/3$ and no heating. We consider the
coolings to be due to inverse bremsstrahlung, and the transfer of energy
to the electrons by Coulomb coupling. 
In the equation which governs the electron energy, 
we use heating due to transfer of energy from protons through
Coulomb coupling and cooling due to bremsstrahlung and Comptonization. If the
electron temperature $T_e$ is high enough, $k T_e > m_e c^2$
(where, $m_e$ is the mass of an electron), we use $\gamma\!=\!4/3$, otherwise
we use $\gamma\!=\!5/3$.
Synchrotron losses could be trivially added to our equations. Since 
this requires a knowledge of unknown magnetic field contents, we ignore 
its effects for the time being.  The temperature at the shock is not high 
enough to induce a significant pair production, so we ignore these effects. 
Within the convergent flow regime $r<2-3r_g$, in optically thick flow, pair
production will take place which may be responsible for the annihilation 
lines observed in soft state.

\noindent{\large 2.3 Radiative Processes}

The bremsstrahlung and Coulomb cooling rates are taken from the standard texts
(e.g., Lang 1980). Correct Comptonization rates have been computed 
very recently (T94; TL95) and we discuss them here for the sake of 
completeness. We provide new results on the computation of reflection of the
hard radiation from the cooler disk.

\noindent {2.3.1 Comptonization}

Comptonization is the problem of energy exchange in the scattering of 
photons off the electrons. In the postshock plasma,
as the accretion rate increases and the Thomson opacity,
$$
\tau_T = \int_{r_i}^{r_s} \sigma_T n_e d r ,
\eqno{(8)}
$$
approaches unity, the escaping photons gain energy due to repeated 
scattering with the hot electrons.
For a thermal non-relativistic electron distribution with 
a temperature $T_e$, the average energy exchange per scattering is given 
by ($h\nu, k T_e\ll m_ec^2$),
$$ 
{{<\Delta\nu>}\over{\nu}}~=~{{4kT_e-h\nu}\over{m_ec^2}}.
\eqno{(9)}
$$
When $h\nu \ll kT_e$, photons gain energy due to the Doppler effect
and when $h\nu\gg kT_e$, photons lose energy because of the 
recoil effect. As the radiation passes through the medium, 
the probability of repeated scattering by the same photon decreases 
exponentially although the corresponding gain in energy is exponentially 
higher. A balance of these two factors yields a power law distribution 
of the energy density: 
$$
F_\nu \propto \nu^{-\alpha} .
\eqno{(10)}
$$
In the limit of high energies ($h\nu \gg kT_e$ ) the exponential hard 
tail is formed as a result of the recoil effect (the photons are unable 
to gain more energy than electrons actually have: $h\nu\propto kT_e$).

It is shown in TL95 that the power laws are the exact solutions
of the radiative transfer kinetic equations taking into account the Doppler
effect only. The related spectral indices
are derived for an wide ranges of the optical depths and electron
temperatures of the plasma cloud from the following transcendental
equation as follows:
$$
\alpha=\frac{\beta}{\ln [1+ (\alpha+3)\Theta/(1+\Theta)
+4d_0^{1/\alpha} \Theta^2]} ,
\eqno{(11)}
$$
where
$$
d_0 (\alpha) =\frac{3[(\alpha + 3)\alpha + 4 ] \Gamma (2\alpha + 2)}{(\alpha
 + 3) (\alpha + 2)^2} ,
$$
and $\Theta=k T_e/m_e c^2$ which represents the appropriately
weighted average of the electron temperature in the postshock region
is computed from the model. Here,
$\beta$ is obtained from the eigenvalue $\zeta\!=\!\exp(-\beta)$
of the corresponding eigenfunction for the appropriate radiative
transfer problem. If we approximate the postshock region as a slab,
for  optical depths $\tau_0>0.1$,
$$
\beta= \frac{\pi^2}{12(\tau_0 + 2/3)^2}(1-e^{-1.35 \tau_0})+
0.45 e^{-3.7\tau_0} \ln\frac{10}{3 \tau_0} ,
\eqno{(12a)}
$$
and for very low optical depth $\tau_0 <0.1$,
$$
\beta= \ln\left[\frac{1}{\tau_0 \ln(1.53/\tau_0)}\right] .
\eqno{(12b)}
$$
If on the other hand, the postshock region is approximated
as a spherical bulge, 
then for  optical depths ($\tau_0>0.1$),
$$
\beta= \frac{\pi^2}{3(\tau_0 + 2/3)^2}(1-e^{-0.7\tau_0})+e^{-1.4\tau_0} 
\ln\frac{4}{3\tau_0} ,
\eqno{(13a)}
$$
and for very low optical depth ($\tau_0 <0.1 1$)
$$
\beta=\ln\frac{4}{3\tau_0} .
\eqno{(13b)}
$$ 
In Hua \& Titarchuk (1995, hereafter HT95)
the Comptonization spectra are given for the optically 
thin ($\tau_0<1)$ and thick ($\tau_0>1)$ regimes. 
There is a good approximate formula  combining both regimes 
(HT95 Eq. [9]) which we use in our spectral calculations.
  
From the numerical simulation of the three dimensional flows
(MLC94), it is evident that the postshock region 
behaves more like a thick accretion disk
(Rees et al. 1982; Paczy\'nski \& Wiita 1980). 
Thus, it is neither a slab nor a sphere, but approximately a torus.
Therefore, one expects the spectral index
to be somewhat in between these two limits. However, as one goes inward,
the optical depth rises and the contribution to the external hard radiation
field diminishes. Therefore, either of these two assumptions
should give sufficiently accurate results. The results
we provide in the next section assumes the geometry of the hot region to be
spherical. This assumption is made only to obtain an appropriately
averaged electron temperature to compute the emergent hard spectra.

For our self-consistent calculation of the temperature distribution of the
shock region we use the expression of the Comptonization enhancement factor
$E_{Comp}$ as a function of the spectral index $\alpha$ and the effective 
low-frequency (temperature $T_r$) dimensionless photon energy 
$x_0=2.7kT_r/kT_e$ 
$$
E_{Comp}\!=\!\cases {q_{x_0}(\alpha)x_0^{\alpha-1}, \ \ \ 
& {\rm if \ \ \ \ \ } $\alpha<1$;\cr \displaystyle{\frac{\alpha(\alpha+3)
(1-{{\alpha+4}\over{2\alpha+3}}x_0^{\alpha-1})}
{(\alpha+4)(\alpha-1)}}, & {\rm \ \ \  if \ \ \ \ \ } $\alpha\geq 1$ .\cr}
\eqno(14)
$$
where,
$$
q_{x_0}(\alpha)\!=\!\frac{\alpha(\alpha+3)\Gamma(\alpha+4)
\Gamma(\alpha)\Gamma(1-\alpha)}{\Gamma(2\alpha+4)}(1-x_0^{1-\alpha}).
$$
This formula, obtained by integration of the emergent
spectrum over the photon energy covers asymptotic forms for 
both the regimes ($\alpha < 1$ and $\alpha \geq 1$) which are presented in 
ST80, Sunyaev \& Titarchuk (1985, hereafter ST85). 

\noindent {2.3.2 Computation of the Reflected Component}

Because a great deal of importance has been given in recent literature
on the possibility of observation of the reflected component
(e.g., Zdziarski et al., 1990; Done et al. 1992; Haardt et al., 1993), we
address this issue here in detail. We find angle and energy dependent
analytical solution (for $E<m_e c^2$) which is applicable  when
the reflector (disk)  is cold ($\lsim 10^5$K). For a hotter disk,
the reflection is more efficient because the photoelectric
absorption process is considerably weak (Kallman \& Krolik, 1986).
We have incorporated this effect by neglecting the
photoelectric absorption term (eq. 19-21 below) for $T\gsim 10^5$K.
Our computation of the equivalent width in terms of a cold disk
reflection only gives an  upper limit, since it is smaller for any
hotter disk.

In computation of the reflected hard component, we note that the reflection 
spectrum from the disk is a result of scattering of the hard radiations 
from the shock by the relatively cold matter of the 
disk. They suffer photoelectric absorption at energies less than $10$ 
keV with emission of the strong $K_{\alpha}$ line and 
down-scattering (recoil effect) at energies higher than $50$ keV.
Thus, the reflection spectrum has a maximum at an energy around $15-20$ keV.
In order to estimate the reflected flux one requires a knowledge
of the fraction $f_{sd}$ of the shock hard radiation intercepted by the disk
and the reflection properties of the disk.
In the very nonrelativistic energy range $\le 30$keV, photons scatter
off electrons almost coherently and consequently the results of the 
classical reflection problem (see e.g., Chandrasekhar 1960) are applicable,
i.e., the intensity reflected from the disk at angle $\theta\!=\!\cos^{-1}{\mu}$
(with respect to the disk normal) when the incoming hard flux $\pi F_{H}$ 
falls onto the disk at the angle $\theta_0\!=cos^{-1} {\mu_0}$ is
as follows
$$ 
I(\mu,\mu_0)={{\lambda_{\nu}F_H f_{sd}}\over{4}}
{{H(\mu, \lambda_{\nu})H(\mu_0, \lambda_{\nu})}\over{\mu+\mu_0}} .
\eqno{(15)}
$$
Here, 
$$
\lambda_{\nu}\!=\{[1+[Y+Y_0\cdot\Theta(E-7.11 {\rm keV})]
(7.8~{\rm keV}/E)^3\}^{-1}, 
\eqno{(16)}
$$
is the photon scattering probability in the presence of photoelectric
absorption, $H(\mu, \lambda_{\nu})$ is the well known H-function 
which changes within a range of $1-3$ or even less depending on 
the photon scattering probability (see e.g., Chandrasekhar, 1960),
$Y$ and $Y_0$ are the abundances of elements (in units of the cosmic 
abundance) with a charge $Z<26$ and the iron abundance, respectively. 
$\Theta(x)$ is the Theta-function 
($0$ for $x<0$ and $1$ for $x>0$). At the hard tail, 
the intensity can be presented as a product of the three
factors, namely $I(\mu,\mu_0)\!=\!F_H f_{sd} {\cal A}_{\nu}(\mu,\mu_0)$,
where $A_{\nu}(\mu,\mu_0)$ is the albedo of the cooler disk.

\noindent {\it Computation of Albedo}

In the computation of the albedo ${\cal A}_\nu (r,\mu,\mu_0)$ of the disk, 
same two processes are to be taken into account: the recoil effect and 
the photoelectric absorption. In the presence of the recoil
effect alone, the incident radiation of frequency $\nu$ will be completely
absorbed, if (ST80, Titarchuk 1987, hereafter T87)
$$
\frac{\Delta \nu}{\nu} \sim \frac{h\nu}{m_e c^2} \tau_0^2 = z \tau_0^2
\sim 1
\eqno{(17)}
$$
i.e.,
$$
\tau_0 \sim \frac{1}{z^{1/2}}.
\eqno{(18)}
$$
The albedo ${\cal A}_\nu$ would be $1-1/\tau_0\!=
\!1-\sqrt{z}$. In the presence of photoelectric absorption, similarly 
{$\tau_0^2(1-\lambda_{\nu})\sim \!1$ and ${\cal A}_\nu\!=
\!1-1/\tau_0 \! =\! 1-\sqrt{\delta_\nu}$, where,
$$
\delta_{\nu}\!=\!1-\!\lambda_{\nu}\! 
=\!\frac{[Y+Y_0\cdot\Theta(E-7.11 {\rm keV})] (z_*/z)^3}
{1+[Y+Y_0\cdot\Theta(E-7.11 {\rm keV})](z_*/z)^3}
$$
($z_*\!=\!7.8~{\rm keV}/511{\rm keV}$) is the probability of the 
photoelectric absorption (see, Eq.16).
\par
In the presence of
both recoil and photoelectric absorption, and taking into account 
the angular dependence of the incoming and outgoing radiations, we 
(see T87) obtain a
more accurate  expression for albedo as a function of the frequency of the
incoming radiation and the angle of incidence $\theta$ as,
$$
{\cal A}_\nu=1-\phi(\mu_0)\Delta ,
\eqno{(19)}
$$
where
$$
\phi(\mu_0)=\frac{1+\sqrt{3} \mu_0}{1+\sqrt{3(1-\lambda_\nu)}\mu_0}
[1-\frac{\lambda_\nu \mu_0 }{4}(1+\lambda_\nu^2)(\ln \mu_0 + 1.33-1.458 
\mu_0^{0.62})],
\eqno{(20)}
$$
$$ 
\Delta=(1-\lambda_\nu)^{1/2} 
=[\sqrt{z} {\rm exp} (\frac{\delta_\nu}{z}) {\rm erfc \ }
(\frac{\delta_\nu^{1/2}}{\sqrt{z}}) + \delta_\nu^{1/2}],
\eqno{(21)}
$$
and $\mu\!=\!\cos \psi$ where $\psi$ is the angle which the  incident
radiation subtends with the local normal to the disk. The second
term is negligible for $T \gsim 10^5$K.

\noindent {2.3.2.2 The analytical solution of the reflection component}

The problem of the reflection from the cold material  
is related to the radiative transfer kinetic
equation which takes into account the photoelectric absorption and the 
recoil effect (see e.g. Basko, Sunyaev \& Titarchuk 1974, Illarionov et al.
1979, T87, Grebenev \& Sunyaev 1987, Magdziarz \& Zdziarski 1995).
We take into account contribution of the multiple scattering, using 
Fokker-Planck (diffusion) approximation and the first scattering exactly.
This is a standard approach used in the classical radiative transfer theory 
for calculation of the coherent multiple scattering (e.g. Sobolev 1975).  

The illumination of the semi-infinite atmosphere (disk) by the parallel 
flux $\pi F(z)$ with inclination angle with respect to the disk normal
$\theta_0=cos^{-1}{\mu_0}$ produce the exponential primary photon distribution
throughout the atmosphere as follows
$$
f(\tau,z)= \frac{\lambda_{\nu}F(z)}{4}\exp({-\tau/\mu_0}),
\eqno{(22)}
$$   
where $\tau$ is the Thomson optical depth calculated from the upper
disk surface.
The first scattering component of  the emergent spectral intensity
observed at the  angle 
$\theta =cos^{-1}{\mu}$ with respect to the disk normal is calculated by 
the integration of the primary photon distribution with 
the escape probability $\exp(-\tau/\mu)/\mu$ along the line of sight
(see, e.g., Chandrasekhar 1960)
$$I_1(\mu,\mu_0,z)=\frac{\lambda_{\nu}F(z)}{4}\frac{\mu_0}{\mu+\mu_0}
\eqno{(23)}
$$   
The multiple scattering component is produced by the integration of
the Fokker-Planck (Kompaneets) equation  and by  integration of
the product of its solution and 
the escape probability $\exp(-\tau/\mu)/\mu$ along the line of sight.

The Fokker-Planck (Kompaneets) equation reads (T87, GS87)
$$\frac{1}{z^2}\frac{\partial}{\partial z}(\eta(z)z^2 n) +
\frac{1}{3\phi(z)}\frac{\partial^2 n}{\partial\tau^2} - 
\frac{\sigma_A}{\sigma_T}n=-\frac{\lambda_{\nu}F(z)}{4}\exp(-\tau/\mu_0).
\eqno{(24)}
$$
where $n(z,\tau)$ is the photon occupation number,
$$\frac{\sigma_A}{\sigma_T}=1-\lambda_{\nu}\approx
[Y+Y_0\cdot\Theta(E-7.11 {\rm keV})](7.8~{\rm keV}/h\nu)^3,$$  
$$\phi(z)=(1+2.8z-0.44z^2)^{-1},
\eqno{(25)}
$$
$$ \eta(z)=\frac{z^2}{1+4.6z+1.1z^2}.
\eqno{(26)}
$$
The solution of this equation with the appropriate boundary condition
at $\tau_0=0$ (${\partial n}/{\partial \tau}-3/2=0$) is given by the integral
(T87, GS87)
$$J(\tau,z,\mu_0)=\frac{z}{4\eta(z)}\int_x^\infty \exp\left(-\int_z^{z_{1}}
\frac{\sigma_A(t)}{\eta(t)\sigma_T}dt\right)$$
$$\cdot \frac{\lambda_{\nu}F(z_1)}{z_1}dz_1
\int_0^{\infty}\exp(-\xi/\mu_0)P[\tau, u(z,z_1),\xi]d\xi.
\eqno{(27)}
$$
where
$J(\tau,z,\mu_0)=z^3 n(\tau,z,\mu_0)$ is the average intensity,
$$
P(\tau,u,\xi)=\frac{\sqrt 3}{2\sqrt{\pi u}} \{\exp[-3(\tau-\xi)^2/4u]
+\exp[-3(\tau+\xi)^2/4u]$$
$$-3\int_0^\infty\exp[-3(\tau+\xi+\eta)^2/4u-
3\eta/2]d\eta\},
\eqno{(28)}
$$
$$u(z,z_1)=\frac{1}{z}-\frac{1}{z_1}+7.4\ln (z_1/z)+13.54(z-z_1),
\eqno{(29)}
$$
$$\int_z^{z_{1}}\frac{\sigma_A(t)}{\eta(t)\sigma_T}dt=
3.375\cdot10^{-6}[Y+Y_0\cdot\Theta(E-7.11 {\rm keV})]$$
$$\cdot[0.25(z^{-4}-z_1^{-4})+1.53(z^{-3}-z_1^{-3})+0.55
(z^{-2}-z_1^{-2})].
\eqno{(30)}
$$
Evaluation of the internal integral in equation (27) over $\xi$ followed
by the integration (over $\tau$) of the product of the average intensity 
$J(\tau,z,\mu_0)$ and
the escape probability $\exp(-\tau/\mu)/\mu$ along the line of sight gives us
the multiple scattering component of the emergent spectral intensity
$$I_m(\mu,\mu_0,z)=\frac{z}{4\eta(z)}\int_x^\infty\exp\left(-\int_z^{z_{1}}
\frac{\sigma_A(t)}{\eta(t)\sigma_T}dt\right)$$
$$\frac{\lambda_{\nu}F(z_1)}{z_1}
G[\mu, \mu_0,u(z,z_1)/3]dz_1.
\eqno{(31)}
$$
where
$$G(\mu,\mu_0,t)=\frac{\mu_0}{(\mu^2-\mu_0^2)(\mu_0-2/3)(\mu-2/3)}
[\mu_0(\mu^2-4/9)\varphi(t,\mu_0)$$
$$-\mu(\mu_0^2-4/9)\varphi(t,\mu)+
2/3(\mu_0^2-\mu^2)\varphi(t,2/3)],
\eqno{(32)}$$
where $t=u/3$ and
$$\varphi(t,\mu)\!=\!{\rm erfc}(\sqrt t/\mu)\exp(t/\mu^2).
\eqno{(33)}
$$
Thus, the final expression of the emergent spectral intensity 
is a sum of two components (eqs. [23] and [31])
$$
I(\mu,\mu_0,z)=I_1(\mu,\mu_0,z)+ I_m(\mu,\mu_0,z).
\eqno{(34)}
$$  

\noindent {2.3.3 Spectral Properties of the Converging Inflow}
 
As the optical depth of the flow becomes larger than $1$, the momentum
deposition by the bulk motion on the cooler soft photons becomes 
important(e.g. Blandford \& Payne 1981; Lyubarskii \& Sunyaev, 1982). In 
the soft state, when the postshock region becomes cool and optically thick,
the spectral properties would be dictated by this converging inflow.
 
The parameter which shows the importance of the bulk Comptonization
effects relative to thermal Comptonization is $\delta=1/\Theta\dot m$.
For $kT_e\ll m_ec^2$ and $\dot m > 1$, $\delta\gg 1$ and this leads
to an asymptotic relation for the spectral index $\alpha$
$$
\alpha=2\lambda^2-3 .
$$
Therefore, as long as $\delta\gg 1$, bulk Comptonization
effects dominate and the values of $\alpha$ are indepedent
of the temperature of the electrons. For details of derivations of
these parameters, see TMK96.
 
\noindent {2.3.3.1 Spectral Index of Comptonization by Converging Inflow}

In order to understand the origin of the hard component in converging
inflow, one requires to solve an equation of the spectral energy flux
$F(r,\nu)$ with the boundary conditions of $F\rightarrow 0$ at $r
\rightarrow \infty$, and $F=-0.5 n x^3$ at the inner boundary: $r \rightarrow1$ (half the flux is absorbed by the hole).
This boundary condition along with the radiative transfer equation  imply
that eigenvalue $\lambda^2$ is the root of the equation
$$
\left({5\over2}-{{2\lambda^2-3/2}\over3}\tau_b\right)
\Phi(-\lambda^2+5/2,7/2,\tau_b)
+ {{5-2\lambda^2}\over7}\tau_b\Phi(-\lambda^2+7/2,9/2,\tau_b)=0,
$$
where $\tau_b=1.5\dot m$ and $\Phi$ is the well known Kummer's function
(e.g. Abramowicz \& Stegun, 1964).
In the limit of the large $\dot m$, the first root of this equation
$\lambda_1^2=\lambda^2\approx 2.25$ (TMK96) and thus $\alpha=1.5$.
Though this slope is obtained from Newtonian considerations,
similar result is likely to remain valid even in general relativistic
computation. This will be discussed elsewhere.

\noindent{2.3.3.2 Luminosity and Effectiveness of Comptonization 
through Converging inflow}
 
The relation between the luminosity of the low frequency
photon source $L_0=\int_0^\infty \delta(x-x_0)dx=1$ and that of the source
of hard radiation photons subjected to the converging inflow Comptonization
($x=h\nu/kT_e$),
$$
L=\int_0^\infty I(x)dx
\eqno(35)
$$
is obtained using the
integration technique developed in ST80, LS82, ST85,
$$E_{Comp}=\displaystyle{{L}\over{L_0}}=
\displaystyle{\alpha(\xi-1)}
 \cases{\displaystyle{{{1}\over {\xi(\alpha-1)}}
\left(1- {{\xi}\over{\xi+\alpha-1}}
x_0^{\alpha-1}\right)},
~~&{{\rm for}~~$\alpha \geq 1$} \cr
\displaystyle{{\Gamma(\xi)\Gamma(\alpha)
\Gamma(1-\alpha)(1-x_0^{1-\alpha})x_0^{\alpha-1}}\over
{\Gamma(\alpha+\xi)}},~~ &{{\rm for}~~$\alpha\leq 1$}}
\eqno (36)$$
where $\xi=\alpha+4+\delta$.
 
This is a generalization of ST80 results for the converging inflow
case. Eq. (36) also has a continuous transition through spectral index
$\alpha=1$ (see eq. 14).
 
The important conclusion from the above exercise is that the low-frequency
source flux is amplified only by a factor of 3 ($\alpha\sim 1.5$)  due to
Comptonization of the converging inflow into black hole. Thus the
component is very weak and mainly visible in the soft state when either the
postshock region is cold or when the shock is absent.

\noindent{\large 2.4 Solution Procedure}

A fully self-consistent solution of the system shown in Fig. 1
involves the following considerations:

\noindent (1) The postshock region is hotter and thicker: $h(r_s) \sim 
a_+(r_s) r_s^{3/2}$, $a_+$ being the sound speed in the postshock region. 
It allows this region to intercept a fraction $f_{ds}$ of
the soft photons from the standard disk component. We compute
self-consistently this intercepted soft component from each radial 
distance of the disk and integrate over the disk to obtain $f_{ds}$. 
This intercepted radiation is Comptonized in the postshock disk.
From the two-temperature equations (eq. 6) we obtain
electron temperature as a function of the optical depth. We then
obtain an average electron temperature and the enhancement factor
following the prescription of ST85 and Titarchuk (1988, hereafter T88).
In reality, we approximate the numerical averaging curves
of ST85 and T88 by the following accurate analytical function
(valid for spherical geometry):
$$
g(\tau)=(1-\frac{3}{2}e^{-(\tau_0+2)}) cos \frac{\pi}{2}(1-\frac{\tau}{\tau_0})
+\frac{3}{2} e^{-(\tau_0+2)}
$$
and obtain the average electron temperature from,
$$
T_e(\tau_0)=\frac{\int_0^{\tau_0} T_e(\tau) g^2(\tau) (\tau_0-\tau)^2 d\tau}
{\int_0^{\tau_0} g^2(\tau) (\tau_0-\tau)^2 d\tau}
$$
Here, $\tau_0$ is the total optical depth of the postshock region.

\noindent (2) After the computation of the Comptonization process, we 
calculate the fraction $f_{sd}$ of this hard radiation that is intercepted
by the standard disk. The interception is computed at each
radial distance of the disk and then integrated over the whole disk.
The rest ($1-f_{sd}$) of the hard radiation is directly
radiated away to observers at infinity. 

\noindent (3) We compute the albedo ${\cal A}_\nu (r)$ of 
the preshock standard disk which determines the fraction of the intercepted
flux scattered away at each radius of the disk. 
The rest ${\cal B}_\nu(r)\! =\! 1-{\cal A}_{\nu} (r)$ is assumed to be 
absorbed by the disk and is reradiated.

\noindent (4) In the zeroth order calculation, we assume that the
equatorial Keplerian disk emits exactly as a standard Shakura-Sunyaev
disk for $r>r_s$. After it absorbs a fraction $f_{sd} {\cal B}_\nu (r)$
of the hard-radiation and reprocesses it, the temperature of the
disk is computed at each radial distance.

We now give a simple argument why our algorithm (1-4) should converge:
Let $L_{I}\!=\!f_{ds} L_{SS}$ denote the fraction of
the Shakura-Sunyaev (1973) disk luminosity $L_{SS}$ of the soft radiation 
intercepted by the bulge of the shock, 
$E_{Comp}\!=\!L_{H}/L_I$ denote the enhancement factor of this radiation
due to cooling of the electrons through inverse Comptonization,
and $f_{sd}$ ($\sim 0.25$ for a spherical bulge) denote the
fraction of $L_{H}$ intercepted back by the disk.
The soft component observed from a disk around the black hole candidate is 
given by equation (2).
The enhancement factor $E_{Comp} \sim 3(T_{e}/3T_{d})^{1-\alpha} \sim 10-30$
(for $\alpha \sim 0.7-0.8$) because, typically,
the electron temperature $T_{e} \sim 50$ keV and the disk temperature
$T_d \sim 5$ eV for parameters of active galaxies and $T_{e} \sim 150$ keV
and $T_d \sim 100$ eV for stellar black hole candidates. Hence, we easily 
achieve a convergence, $f_{ds} E_{Comp} f_{sd} {\cal B}_\nu < 1$ 
since $f_{sd}\sim 0.25$, $f_{ds} \sim 0.05$, and ${\cal B}_\nu \sim 0.5$.

In each run, we repeat steps (1) through (4) until the 
temperature distribution and the spectral index $\alpha$ converges. 
Typically,  $5$ to $10$ iterations are enough to achieve convergence.

\noindent{\Large 3. RESULTS AND INTERPRETATIONS}

In each case we compute, we choose a set of four quantities, namely, the mass
of the central black hole $M$, the shock location $r_s$
(instead of angular momentum of the halo), the accretion
rate of the disk ${\dot m}_d\!=\!{\dot M}_d/{\dot M}_{Edd}$ and the 
accretion rate of the halo ${\dot m}_h\!=\!{\dot M}_h/{\dot M}_{Edd}$. We shall
present a series of runs for a galactic black hole candidate, 
$M\!=\!5M_\odot$, and a series of runs when a black hole is massive, 
$M\!=\!10^7M_\odot$. In both the cases, we shall consider parameters
suitable for a non-rotating black hole ($r_s\!=\!10r_g$, $r_i\!=\!3r_g$).
By choosing two independent accretion rates, we thus do away with
the unknown viscosity parameter.

In Fig. 2, we show the temperature distribution inside the disk
and the postshock region when ${\dot m}_d\!=\!0.001$ (solid line), $0.01$ 
(long-dashed line), $0.1$ (short-dashed line) and $1$ (dotted line), 
${\dot m}_h \!=\!1$ is chosen throughout. We choose $M\!=\!5M_\odot$.
As the accretion rate of the Keplerian disk is increased, its temperature
is increased. The postshock region, on the other hand, is
{\it cooled}, since the number of soft photons intercepted by the
postshock region is increased. The more the electrons cool through
the inverse Compton process, the more the protons supply them energy through
Coulomb coupling. Eventually, for ${\dot m}_d\!=\!1$, the {\it runaway} process
is seen where both the electrons and protons become cold catastrophically.
Subsequently, the cooler but denser flow cools through the bremsstrahlung
process alone. Exactly similar behavior is seen when the computation
around a massive black hole is also carried out.

If Comptonization is the dominant mechanism for cooling and the
densities and velocities vary as in a spherical flow, i.e.,
$v \propto r^{-\frac{1}{2}}$ and $\rho \propto r^{-\frac{3}{2}}$,
one can obtain an analytical expression for the electron temperature 
from the following equation (see, eq. [6]):
$$
\frac{1}{T_e}\frac{dT_e}{dr} +\frac{1}{r}-C_{Comp} r^{1/2} \sim 0 ,
\eqno{(37)}
$$
where $C_{Comp}$ (obtainable easily from eq. [6]) is a monotonically
increasing function of 
optical depth and the spectral index is assumed to be constant for a 
given case. The solution of the above equation is
$$
T_e=\frac{T_{es}r_s}{r} e^{C_{Comp}(r^{3/2}-r_s^{3/2})} .
\eqno{(38)}
$$
This shows that if $C_{Comp} > r_s^{-3/2}$, the cooling due to
Comptonization overcomes compressional (geometric) heating
and $T_e$ drops as the flow approaches the black hole,
and the convergent flow regime begins.
Our numerical integration of equation (6) requires a somewhat higher
amount of soft photons for cooling because of other heating effects,
such as energy gain from the protons due to Coulomb process.
The catastrophic cooling of the postshock region in soft states
in important, since it introduces seed soft photons within the `convergernt
flow' regime and produce the weaker hard  component of slope $\sim 1.5$.

In Figure 3a, we present the variation of the
energy spectral index $\alpha$ (observed slope in the $2-50$keV region) 
when a black hole of mass $M\!=\!5 M_\odot$ is chosen. The
abscissa is the logarithmic mass accretion rate of the Keplerian disk 
component (${\dot m}_d$) and the different curves are marked by
the mass accretion rates of the halo component (${\dot m}_h$).
We compared the results from the postshock flow as well as
the convergent inflow. The dashed curves are drawn where
the spectra could be contaminated by both effects.
In Fig. 3b, we show mean electron temperature.
The solid curve is for ${\dot m}_h\!=\!1.0$ and the long-dashed and 
short-dashed curves are for ${\dot m}_h\!=\!2.0$ and $0.5$ respectively.
For a given halo rate,
a transition between states can be achieved by a change in ${{\dot m}_d}$.
Qualitatively, the cooling rate due to Comptonization 
for unit optical depth is proportional to the electron 
temperature and the local energy density,
i.e., $L_{H} / \tau_h \propto T_e (L_{H} + L_{I})$. 
Hence, $\tau_h T_e \propto (1+L_{I}/L_{H})^{-1}$. 
In the hard state, $L_{I} \ll L_{H}$ and $\tau_h T_e \sim $ constant. 
When $\tau_h\lsim 1$, this leads to $\alpha \sim$ constant 
(T94, TL95, see also, Pietrini \& Krolik, 1995).
This is what we observe as well: $\alpha$ is insensitive 
to the ${\dot m}_d$ when ${\dot m}_d \lsim 0.1$. In the
soft state, $L_{I}>>L_{H}$ and $T_e \sim \frac{L_{H}}
{L_{I}}\frac{1}{\tau_h}$. Smaller $T_e$ leads to a very high $\alpha$,
as observed. For a smaller ${\dot m}_h$, the optical depth $\tau_h$ becomes
small and the temperature high, causing
the index to become especially sensitive to both ${\dot m}_d$ and 
${\dot m}_h$. When accretion rate of the disk is very high, the spectral
index asymptotically becomes $1.5$, as is observed in many systems
(e.g. Sunyaev et al. 1994, 
Miyamoto et al. 1991, Parmar et al. 1993, Wilson \& Rothschild
1983). Successful fits of soft spectra of black hole
candidates  GS1124-68 (Ebisawa et al., submitted) and GRS1009-45
(Titarchuk et al., submitted) have been obtained using our model.
These fits are therefore consistent with a freely-falling,
sub-Keplerian component close to a black hole.

In Figures 4(a-b), we present the same quantities computed 
for a massive black hole ($M\!=\!10^7M_\odot$) candidate.
The results are similar, though the average temperature is smaller
and the index $\alpha$ is more sensitive to
the disk rate when ${\dot m}_d \gsim 1$ and ${\dot m}_h \gsim 1$. 
Here, the rapid rise of $\alpha$ is due to the abrupt cooling (see Fig. 2)
of the postshock region within a narrow layer.

Figure 5 shows the contribution of the component spectra for a particular
case with ${\dot m}_d\!=\!0.1$, ${\dot m}_h\!=\!1.0$, $M=5M_\odot$.
For illustration, the observation angle $\theta$ is
chosen so that $\mu\!=\!cos\theta\!=\!0.4$. The long dashed curve
represents the soft spectrum from the Keplerian disk, while the short dashed
curve is the contribution to the soft spectrum from the reprocessed
and intercepted hard radiation by the Keplerian disk (second
term in eq. [2]). The dotted curve is the contribution to the
hard radiation from the postshock region. The dash-dotted curve 
denotes the reflected hard radiation from the Keplerian disk.
The sum of these contributions is depicted by the solid curve.

Figure 6 shows a comparison of the four runs for the spectra around
a black hole of mass $5 M_\odot$ and ${\dot m}_h\!=\!1.0$. 
The disk accretion rates (${\dot m}_d$)
are $0.001$ (solid line), $0.01$ (long-dashed line), $0.1$ 
(short-dashed line) and
$1.0$ (dotted line) respectively. As shown in Figs. 3a, 3b, 4a, 4b,
with the increase of the disk accretion rate, the temperature 
$T_{e}$ of the electrons is reduced and the energy index $\alpha$
is increased. The luminosity and the peak frequency of the soft component
go up monotonically with ${\dot m}_d$. The hard component shows a `pivoting' 
property: the intensity of the hard component rises with ${\dot m}_d$ 
below $\nu \sim 10^{18-19}$ Hz (corresponding energy $\sim 5-50$ keV) 
but it falls at higher energies since the electrons become cooler.
In the soft state, the postshock
region becomes cooler and produces a weak hard component of slope
($\alpha \sim 1.5$) due to the convergent flow
as shown by the dash-dotted curve. In Fig. 7 we show
this `power-law' feature with the components separated. The long-dashed
curve is the soft-luminosity from the Keplerian accretion disk and
the short-dashed curve is due to the Comptonization
of the intercepted ($\sim 5\%$) soft photons in the convergent flow.
This power-law component is possibly the
only true signature of black hole and is observed
in black hole candidate spectra (e.g. Sunyaev et al. 1994).
In the neutron star case, the flow need not pass through the
inner sonic point (C89, C90a,b) if it is not compact enough and the flow
is subsonic at the surface. Thus, no significant bulk momentum is added
to the soft photons to create this hard component.
Indeed, in this case, turbulent mixing  of halo and disk components
in the postshock region becomes important,
which results in a hard component of slope $< 1$ and
electron temperature $T_e \sim 12$ keV which is
observed (Claret et al. 1994).

If the angular momentum of the overall disk, or
even the sub-Keplerian component, is quite low, $l \ll l_{ms}$,
the disk need not have a shock at all. Shock free advective
disk (orginating out of Keplerian component, see, Appendix)
can also intercept soft photons from the Keplerian disk and show
similar behaviors mentioned here. It is unclear whether
X-ray emitted in this process would be efficient. 
As the viscosity of the disk varies, the sub-Keplerian flow 
would meet the Keplerian disk at varying distance (Appendix, C95b)
and could mimic the hard-to-soft state transition. However,
in this case soft and hard components would always be correlated
which is not observed.

Increasing the halo rate ${\dot m}_h$ increases the optical depth of 
the postshock region  thereby increasing the hard component luminosity
$L_{H}$. The spectrum is hardened due to
the saturation of Comptonization (see, Figs. 3a and 3b).
Figures 8a and 8b show the comparison of electron temperatures (in 
K) and the spectra, respectively,
as ${\dot m}_h$ is varied while ${\dot m}_d\!=\!0.1$
is kept fixed throughout. Solid, long-dashed,
and short-dashed curves are drawn for ${\dot m}_h\!=\!1$, $2$ and $0.5$
respectively. Contrary to what is observed in Fig. 6, 
the hardening of the spectra in the present circumstance is 
accompanied by no significant change in the soft luminosity and 
an actual {\it increase} in the hard luminosity.

The above computations use the shock location to be
$r_s \sim 10r_g$ and the inner edge at $3 r_g$. These are
the typical length scales for accretion of matter with 
marginally stable angular momentum (C89, C90a,b). 
An increase in angular momentum of the halo increases the location of 
the shock, cools the average postshock electron temperature and raises
the average spectral index. It also suppresses the soft luminosity 
relative to the hard luminosity. Similarly, a study of the 
behavior in the Kerr geometry with, say, $r_s\!=\!5r_g$ and $r_i\!=\!1.5r_g$
with similar disk and halo parameters indicates a general increase of the
average halo temperature and an average decrease of the spectral index.

The above mentioned properties from our model are summarized
(see also, CTKE95) in Table 1 where the correlation (arrows pointing up) 
or anticorrelation (arrows pointing down)
of the observable quantities (luminosities $L_{X,\gamma}$
in X-ray and $\gamma$-ray regions) with the input accretion rates
are shown. Smaller arrows represent a weaker correlation.

\newpage

\noindent {\large TABLE 1}

\centerline{Variations of the Spectral Properties with Accretion Rates} 

\begin{center}
\begin{tabular}{|l|l|l|l|}
\hline
    & & & \\
\sl  & \sl Input & \sl ${\dot m}_d$ & \sl ${\dot m}_h$ \\
    & {\large $\rightarrow$} & & \\
\hline
Output & Row & & \\
{\large $\downarrow$} &  \&  & \sl (1) &  (2) \\
  & Column & & \\
\hline
    & & & \\
 $L_{S}$ & (1)  & {\Large $\uparrow$} &  {\small $\uparrow$}  \\
    & & & \\
\hline
    & & & \\
$L_X, \alpha$ & (2) & {\Large $\uparrow$}{\Large $\uparrow$}$^a$ 
& {\Large $\uparrow$}{\Large $\downarrow$}$^{b,c}$ \\
    & & & \\
\hline 
    & & & \\
$L_\gamma, \alpha$ & (3) & {\Large $\downarrow$}{\Large 
$\uparrow$}$^a$ & {\Large $\uparrow$}{\Large $\downarrow$}$^b$ \\
    & & & \\
\hline
\end{tabular}
\end{center}
\indent $^a$ dependence is weaker for ${\dot m}_d \lsim 0.1$\\
\indent  $^b$ dependence is weaker for ${\dot m}_h \gsim 1$\\
\indent  $^c$ $\alpha_X \sim 1 \rightarrow 1.5$, $L_X/L_S \leq 10^{-3}$ 
in the convergent flow regime (see, Fig. 3)\\

\noindent{\Large 4. COMPARISON OF THE PREDICTED BEHAVIOR WITH OBSERVATIONS}

So far, we have delineated the spectral properties of the accretion disks
which are formed out of Keplerian and sub-Keplerian matter.
We can now compare these predictions with the observed facts.
Galactic black hole candidates, such as {\it GX339-4}, {\it GS2023+338} and
{\it GS1124-68} show almost constant spectral slopes over two decades
of the luminosity variation. (TAN; Ueda et al. 1994, hereafter U94; E94). 
The constancy of the slope is explained by the lower left corner of Fig. 3a.
The spectral evolution of the black hole candidate novae, 
{\it GS2000+25, GS2023+338}, and {\it GS1124-68} 
are very different (TAN; E94). These differences could 
be accounted for by variations of the properties of 
the halo (Figs. 8a and 8b). For instance, {\it GS2023+338} is always 
in the hard state throughout the outburst, suggesting a 
high ${\dot m}_h >1$ but lower ${\dot m}_d \lsim 0.01-0.1$. The suppression 
of the soft component could also be due to a distant shock. {\it GS2000+25},
on the other hand, remained in the soft state during the outburst, suggesting 
a high ${\dot m}_d \sim 1.0$, but a low ${\dot m}_h \lsim 0.1-0.5$.
In the rising phase of {\it GS1124-68} (E94), 
the increase of $L_{H}$ below $\sim 10$ keV is accompanied by its decrease
above $\sim 10$ keV. This is the `pivoting' property in the hard spectra we
referred to (Fig. 6). This object,
along with other black hole candidates such as {\it Cyg X-1} 
and {\it GX339-4} are observed in two distinctly different states.
Generally speaking, in the hard state,
$\alpha \sim 0.7$ remains almost constant but in the soft state
the hard component is very weak with index $\alpha \sim 1.0-1.5$,
again remaining constant during a the period in which $L_{S}$ varied
by as much as two orders of magnitude (TAN, E93, E94).  
In the soft state, ${\dot m}_d$ is very high and the postshock
optically thick region produces
a spectral index of $\sim 1.5$ as it would in a  
convergent flow (Fig. 6). In {\it GS2023+338}, the tendency of the 
anticorrelation between $L_X$ and $L_\gamma$ is also seen 
(TAN, Sunyaev et al., 1991), but they do not 
actually cross, possibly because both ${\dot m}_d$ and ${\dot m}_h$
change together. The hard component in galactic
black hole candidates is generally found to be more variable than
the soft component and the variations are found to be on time scales
of days to months (TAN, I92, E94). 
This is possibly due to independent variations in accretion rates
as we have chosen in our model.

Among the massive black hole candidates associated with AGNs, {\it NGC 4151} 
has been studied extensively (Yaqoob 1992; Yaqoob et al. 1993). 
{\it NGC 4151} shows $\alpha$ to be generally stable though
it is seen to rise with the hard luminosity in the $2-20$keV
range. This is consistent with our result of variation of ${\dot m}_d$
(Fig. 6) where we show that below the pivoting,
the hard luminosity rises with $\alpha$. The $\alpha$-$L_{H}$ plot
also showed some scatter. This can be understood
from our Fig. 8b, where we show that when the halo rate
is increased, the hard flux increases with the decrease of $\alpha$.
Because of longer time-scales, Seyferts do not show a change of
states. We believe that possibly all Seyfert-1s belong to 
different regions of the same diagram (Figures 4a and 4b). 
A spectral index of $\sim 1.5$ is also reported in MKN841
(Arnaud et al., 1986) whose origin could be the same as that of 
the galactic black holes discussed above.

\noindent{\Large 5. CONCLUDING REMARKS}

In this paper, we have presented the spectral properties of
the accretion flows which are primarily sub-Keplerian
or an admixture of Keplerian and sub-Keplerian at the outer boundary.
Such matter produces an optically thick standard disk on the equatorial plane
and the sub-Keplerian optically thin halo surrounding it produces a shock
close to the black hole or a neutron star. We showed that
the reprocessing of the soft-photons from the standard disk
by the hotter, postshock region of the disk is sufficient
to produce the observed hard radiation with correct behavior
of the luminosity and the spectral slope. We showed that the
the spectral index $\alpha \sim 0.7-0.8$ could remain generally stable
even when the accretion rate of the disk (${\dot m}_d$)
is changed by more than 2 orders of magnitude, as long as ${\dot m}_d 
\lsim 0.1$. This is seen in objects such as, {\it GX339-4, GS2023+338} and
{\it GS1124-68} and possibly in some class of Seyferts. 
When the disk accretion rate is close to the
Eddington rate, the sensitivity of $\alpha$ to the accretion rate
goes up and the object switches to the soft state as the disk accretion rate
is increased. We also showed that during the change of the state, the hard 
luminosity remains the same in some energy band, a property (pivoting) that is 
observed in GS1124-68. Usually, this location of the
pivoting depends upon the accretion rate. A search for this property
in a wide band ($2-200$keV) should be made
in other black hole candidates, such as Cyg X-1 and GX339-4 
to verify our model. A similar property is expected
in X-ray bursters, and other neutron star candidates as well.
We also showed that in the
soft state, the consideration of the convergent flow becomes applicable.
We believe that the spectral slope of $1.5$ that is seen in the
soft state is the signature of a very large converging flow
close to the black hole. For a neutron star accretion,
such a component should clearly be absent. Traces of convergent 
inflow should be observable at very
high energies even in hard state, since the halo rate remains high.

It may be possible that for stellar black holes, the two 
components of the disk, namely, the sub-Keplerian and the Keplerian, 
are of different origin. The sub-Keplerian
component may be from the winds, whereas the Keplerian component
is primarily from the surface of the companion. If they were of same origin
(i.e. from a completely Keplerian or a sub-Keplerian disk),
the variation at the source would be reflected
in the variation of the hard and the soft components at two different times.
This is because the halo is almost freely falling, whereas the disk
is moving on a viscous timescale. Roughly speaking, for a similar
halo and disk thickness variation with radius, the ratio
of the time scales would be $t_{disk}/t_{halo} \sim 1/\alpha_{v}$,
where $\alpha_{v}$ is the Shakura-Sunyaev viscosity parameter
(Shakura \& Sunyaev 1973) inside the disk. However, even this 
correlation may be generally difficult to detect, since,
whereas the variation of the disk rate changes both the soft and the
hard luminosities, the variation of halo rate changes primarily the
hard component. Also, the variation of the hard
flux and the spectral index are inversely correlated with these
rates (Figs. 6 and 8b). Another source of complexity could be the
entrainment and evaporation of some disk matter which joins the halo
and the loss of some matter as winds and outflows. Details of these
processes can be quantitatively understood through numerical simulations
and will be discussed elsewhere.

We have noted above that there could be two sources of strong winds
in the vicinity of the disk. Numerical simulations show that centrifugal 
barrier pushes sub-Keplerian, inviscid flows along the direction of
the jets (Hawley et al. 1984, 1985; 
Eggum, Coroniti, \& Katz, 1987; MLC94, Ryu et al. 1995). Another source of winds
could be the evaporation of matter close to the shock. Judging from the
resemblance of the iron resonance line shapes (Piro, Yamauchi, \& Matsuoka, 
1990; Rajeev et al. 1994; Mushotzky et al. 1995) with the P-Cygni profiles
and the profiles of down-scattered emission lines
we believe that these lines could be generated from the winds themselves.
These profiles are characterized by the 
prominent emission feature at the red wing  and by possibly
weaker absorption feature at the blue wing. 
We are more convinced by the fact that we find the upper limit
of the equivalent width of 
the lines from the disk to be only about a few tens of electron volts, as 
opposed to the observed equivalent width of several hundreds of electron 
volts. A back-of-the-envelope calculation for the ionization parameter 
(Kallman \& McCray 1982) could be done using typical
parameters for the wind. This turns out to be, $\xi\!=\!L_{H}|_{7-10keV}/n_p 
r^2\sim 10^4 (\frac{{\dot m}_h}{{\dot m}_w})_{2} (\frac {v}{c})_{0.1}$
where, $n_p$ is the number density of protons and 
${\dot m}_w$ is the outflow rate of the line driven wind in units of
the Eddington rate. Here,  a tenth of velocity of light 
(subscript $0.1$) and a $50\%$ of loss in halo matter
(subscript $2$) are assumed. The column density $N\!=\!3.5({\dot m}_h)_{1}
(\frac{v}{c})_{0.1}^{-1} (\frac{r_{in}}{20r_g})^{-1}$ g cm$^{-2}$
may be sufficient to produce the observed iron line in the outflow.
The subscript $1$ of ${\dot m}_h$ indicates that one Eddington
rate is assumed. The emergent spectrum of the outflow is a
result of transmission of the injected hard radiation.
Fig. 9 shows the fate of this hard component (with a typical $\alpha
\sim 0.7$ for illustration) when it passes through a wind of $\tau \sim 3$.
The solid curve shows the emitted component from the
postshock region and the long-dashed curve is the component
down-scattered (ST80) by the outflow (for energy change by down scattering,
see, eq. 17). The short-dashed curve
is produced due to the absorption by heavy elements at the
outer layer of the outflow (pronounced iron edge is seen).
We do not include the line emissions.
If the so-called ``reflection bump" originates in the wind, as we suggest,the net effect must be a combination of the solid and the
short-dashed curve weighted appropriately by the covering factor.
In computing the effect of reflection, it is well known
that only a finite layer of an otherwise semi-infinite target participatesin scattering. Thus, the {\it continuum} spectra from
these two approaches (namely, from reflection and transmission)
are indistinguishable. However, the equivalent width of the
emitted lines should be higher in the outflow, because of a
larger (by an order of magnitude or so) covering factor. The
equivalent width of the lines and the strength of our ``transmission bump"should be correlated as they both depend on the covering factor.
In the rotating winds and outflows, splitted emission lines
should be generated. Such observations are reported in
Tanaka (1994) and Tanaka et al. (1995).
High optical depth ($\tau>3$) of the disk components
close to the hole washes out any disk emission features
and thus these lines can only be emitted from the winds.
Down scattering mimics gravitational red-shifts of lines. Thus
the line profiles in these two processes are again
indistinguishable, apart from the sharp cut-off in the red wing
when a disk emission is considered.
 
Combining results presented in Fig. 1, Fig. 3 and Fig. 10, we are now
in a position to construct sequences of a nova outburst and its quiescent
state (e.g., Cannizzo, 1993). In the quiescent state, entire disk has low
viscosity and the Keplerian disk receeds very far away (Fig. 10).
As the viscosity in the equatorial plane rises, Keplerian disk rate
is increased, and the inner edge of the Keplerian component
comes closer to the black hole. This increases soft photons
to be Comptonized, thereby generating an outburst.
This situation is to be compared with an atomic reactor whose energy
output is controlled by insertion of the fuel rods
(here, the Keplerian disk component) by varying degree. 

In our model we have made some approximations, such as the
equations appropriate for conical
geometry (instead of actual torus) to integrate both the
Euler equations as well as the radiative transfer equations.
Also, we assumed a fixed polytropic index rather than 
computing it self-consistently form the heating and cooling 
as the flow accretes. We have
ignored diffusion of hard photons to the preshock region and subsequant
pre-heating of the electrons, as is the characteristics of 
the photo-hydrodynamical shocks (Riffert, 1988; Becker 1988).
However, we do not believe that our basic 
conclusions will be affected by these approximations. 

It is to be noted that our model differs significantly from the
Comptonized soft photon model of two-temperature hot disks
discussed almost two decades ago (Shapiro, Lightman, \& Eardley 1974;
Shakura \& Sunyaev 1976). In these models, the soft and the hard radiations
originate from the same disk and therefore, the variations
of the hard and the soft components should be always correlated,
this does not appear to be observed. There are some other models
in the literature (Wandel \& Liang 1991; Haardt \& Maraschi
1991) which required extra-components and/or extra parameters.
Our present model is somewhat more general than the single component 
model (which included shock waves) of Chakrabarti \& Wiita (1992) 
where only the optical and UV regions were considered. 

If the shock waves do exist, as we assumed in our model, there could be
another manifestation which is observable as well. Recently Molteni,
Sponholz, \& Chakrabarti (1996) have shown that when the accretion 
rates are such that the cooling
time scale roughly matches the infall time scale, the shock
oscillates with a time period comparable to the cooling time. 
Could this be the origin of quasi periodic oscillations (QPOs)
seen in black hole candidates (e.g., Dotani 1992) 
such as GX339-4 and GS1124-68? Possibly, yes.
Particularly noteworthy is the fact that by dynamically moving
back and forth, the shocks modulate the outgoing hard radiation
by as much as $10$ to $15$ percent, whereas the soft radiation
emitted from the preshock disk is only marginally modulated.
This is because of the fractional variation of the emitting area is
larger for the postshock region. These behaviors
have been observed in dwarf novae, such as SS Cygni, VW Hyi, and U Gem
(see, e.g., Mauche, Raymond, \& Mattei, 1995). Unlike the case 
of a dwarf novae outburst, where cooling could be due to 
bremsstrahlung or line effects, in the black hole
candidates, the time scale is governed by Comptonization processes.

Though we have primarily discussed shock waves around a 
galactic or extragalactic black holes,
the entire computation remains unchanged when a neutron star is
considered instead. The typical size of a neutron star ($10$ km) is
under $2.5r_g$ and therefore our computation zone ($3-10r_g$)
is outside that of a neutron star.
For example, the best fit of the hard  spectra of the
burst source MXB 1728-34 by a Comptonization model (ST80) requires
the parameters $kT_e=12$ keV and the optical depth $\tau_0>3.8$
(Claret et al. 1994) which are easily achievable by our model with a
disk and a postshock boundary layer. Though it may be generally
difficult to distinguish the spectral properties of a galactic
Schwarzschild black hole from that of a neutron star, we believe that the
hard component due to the convergent flow (Fig. 6) is possible only in
a black hole accretion since the inner boundary conditions are
completely different (C89, C90a,b).

Works of SKC and LGT are partially supported by the National Research Council
and NASA grant No. NCC5-52, respectively. The authors thank D. Kazanas,
K. Ebisawa, T. Yaqoob and T. Kallman for useful discussions. They also
thank Paul Wiita for carefully reading the manuscript and suggesting
improvements.

\newpage

\centerline {\Large APPENDIX}

The globally complete viscous, isothermal solutions around a
black hole are in C90a,b
where sub-Keplerian flows become Keplerian at large distance.
In this Appendix, we show this even in the presence of
general heating and cooling.
 
We choose a simple, vertically averaged, axisymmetric, polytropic
disk. The equations governing the flow are (C95a,b):
 
\noindent (a) The radial momentum equation:
 
$$
v \frac{dv}{dr} +\frac{1}{\rho}\frac{dP}{dr}
+\frac {l_{Kep}^2-l^2}{r^3}=0,
\eqno{(A.1a)}
$$
 
\noindent (b) The continuity equation:
 
$$
\frac{d}{dr} ( \rho r h v) =0 ,
\eqno{(A.1b)}
$$
 
\noindent (c) The azimuthal momentum equation:
 
$$
v\frac{d l(r)}{dr} -\frac {1}{\rho r h}\frac{d}{dr}
(\frac{\alpha P r^3 h} {\Omega_{Kep}} \frac{d\Omega}{dr}) =0
\eqno{(A.1c)}
$$
 
\noindent (d) The entropy equation:
 
$$
\Sigma v T \frac{ds}{dr} = Q^+ - Q^-
\eqno{(A.1d)}
$$
These equations are the generalization of viscous, transonic, isothermal
disks for which the global solutions are already in the literature (C90a,b).
Here $l_{Kep}$ and $\Omega_{Kep}$ are the Keplerian angular momentum and
Keplerian angular velocity respectively, $\Sigma$ is the  density $\rho$
vertically integrated, $P$ is the total pressure, $h$ is the vertical
thickness of the disk at radial distance $r$, $v$ is the radial
velocity, $s$ is the entropy density of the flow, $Q^+$ and $Q^-$
are the heat gained and lost by the flow. We compute $h(r)$
assuming the disk is in hydrostatic balance equation in vertical direction.
$l(r)$ is the angular momentum distribution of the disk matter.
Here, we have chosen geometric units, thus $r$ is units of
Schwarzschild radius $r_g=2GM/c^2$, $l(r)$ is in units of $2GM/c$,
and velocities are in units of the velocity of light $c$, $M$ being the
mass of the central black hole. To mimic the black hole geometry
(i.e., to compute the Keplerian quantities above) we use Paczy\'nski-Wiita
(1980) potential. In eq. (A.1c), $\alpha_v\lsim 1$ in the above equation
is the viscosity parameter of
Shakura \& Sunyaev (1973), which is widely used to describe the viscous
stress: $w_{r\phi}=- \alpha P$. Eqn. A.1c could be easily integrated
to obtain the disk angular momentum distribution.  $l(r)-l(r_{in})
=\alpha_v r a^2 / v$ (C90a,b). Without any loss of generality,
we assume $Q_-$ to be a fraction of $Q_+$ while integrating above equations.
Also, in order that the angular momentum remains continuous
across shock waves, we choose total pressure
(thermal plus ram) in the viscosity prescription (CM95). One needs to supply
five quantities (over-determined system, since we are not explicitly
computing the heating and coolings to illustrate our point here): the
angular momentum at the inner edge $l(r_{in})$, the location of the
inner sonic point $r_{c}$, the viscosity parameter $\alpha_v$, the
polytropic index $\gamma=4/3$ and the cooling law $Q_-/Q_+$.
 
Fig. A.1 shows examples of the ratio of the disk angular momentum
distribution to the Keplerian distribution (for $r<3r_g$,
we keep $l_{Kep}(r) = l_{Kep} (3r_g)$ for stability reasons)
as a function of distance from the black hole.
The ratios show deviation from Keplerian due to advection, pressure and
viscous effects. Three diverse cases have been chosen to illustrate
our points. In Case A (marked `A'), we choose $l(r_{in})=1.88$, $r_c=2.2$,
$\alpha_v=0.005$, $\gamma=4/3$ (radiation pressure dominated)
and $Q_-=Q_+$. The flow deviated from Keplerian disk at $7.5r_g$ and
even becomes super-Keplerian (ratio $>1$) close to the black hole.
In Case B (marked `B'), we choose $r_{c}=2.3$, $l(r_{in})=1.7$,
$\gamma=5/3$ (gas pressure dominated), $\alpha_v=0.02$ and $Q_-=Q_+$.
Here the flow deviates from Keplerian at $90r_g$ and always remained
sub-Keplerian. In both the cases above, there is no shock formation.
In Case C (marked `C'), the flow passes through a shock at $13.9r_g$
(angular momentum distribution remained continuous) and remains
completely sub-Keplerian after deviating from the Keplerian disk
at $480r_g$. The parameters chosen are $l(r_{in})=1.6$, $r_{c}=2.87$,
$\gamma=4/3$, $\alpha_v=0.05$ and $Q_-=0.5 Q_+$.
It is easily shown that a decrease of viscosity, keeping
other parameters fixed, increases the distance where Keplerian
disk begins. This property is crucial in understanding our model
of the generalized disk (Fig. 1) and the nova outbursts.

\newpage

\centerline{REFERENCES}

\noindent Abramowitz, M., \& Stegun, I.A. 1964,
Handbook of Mathematical Functions, U.S. Dept. of Commerce, Washington. D.C.\\
Abramowicz, M., Czerny, B., Lasota, J.P., \& Szuskiewicz, E., 1988,
ApJ, 332, 646\\
Arnaud, K. et al. 1985, MNRAS, 217, 105
Basko, M.M. 1978, ApJ, 223, 268\\
Basko, M.M., Sunyaev, R.A., \& Titarchuk, L.G. 1974 A\&A, 31, 249\\
Becker, P., 1988, ApJ, 327, 772\\
Blandford, R.D. \& Payne, D.G. 1981, MNRAS, 194, 1033\\
Cannizzo, J.K., 1993 in {\it Accretion Disks in Compact Stellar
Systems}, ed. J. Craig Wheeler (Singapore: World Scientific), 6.
Chakrabarti, S.K. 1989, ApJ, 347, 365 (C89)\\
Chakrabarti, S.K. 1990a, MNRAS, 243, 610 (C90a)\\
Chakrabarti, S.K. 1990b, Theory of Transonic Astrophysical Flows 
(Singapore: World Scientific, Singapore, 1990) (C90b)\\
Chakrabarti, S.K. 1993, Numerical Simulations in Astrophysics, 
Eds. J. Franco et al. (Cambridge Univ. Press: Cambridge) (C93)\\
Chakrabarti, S.K. 1994, in Proceedings of the 17th Texas symposium,
(New York Academy of Sciences, New York) \\
Chakrabarti, S.K. 1995a, in Accretion Processes on Black Holes,
Physics Reports (in press) (C95a)\\
Chakrabarti, S.K. 1995b, ApJ Letters, submitted, (C95b)\\
Chakrabarti, S.K., \& Molteni, D. 1995,  MNRAS, 272, 80 (CM95) \\
Chakrabarti, S.K., Titarchuk, L., Kazanas, D., \& Ebisawa, K. 1995, 
A\&A Suppl. Ser., Proceedings of 3rd Compton Symposium (CTKE95)\\
Chakrabarti, S.K., \& Wiita, P.J. 1992, ApJ, 387, L21\\
Chandrasekhar, S. 1960, Radiative Transfer, (New York: Dover)\\
Chen, X. M., \& Taam, R. 1993, ApJ, 412, 254\\
Chen, X. M., Abramowicz, M., Lasota, J.P., Narayan, R. \& Yi, I.
1995, ApJ, 443, L61\\ 
Clavel, J. et al. 1990, MNRAS, 246, 668\\
Claret, A. et al. 1994, ApJ, 423, 436\\
Done, C. et al. 1992, ApJ, 395, 375\\
Dotani, Y, 1992 in Frontiers in X-ray Astronomy, (Tokyo: Universal
Academy Press ), 152\\
Ebisawa, K. et al. 1993, ApJ, 403, 684 (E93)\\
Ebisawa, K. et al. 1994, PASJ, 46, 375 (E94)\\
Eggum, G.E., Coroniti, F.V., \& Katz, J.I. 1987, ApJ, 323, 634\\
Grebenev, S.A., \& Sunyaev, R.A. 1987, Sov. Astron. Lett. 13, 438\\
Haardt, F. et al., 1993, 411, L95\\
Haardt, F., \& Maraschi, L., 1991, ApJ, 380, L51\\
Hawley, J.F., Smarr, L.L., \& Wilson, J.R. 1984, ApJ, 277, 296\\
Hawley, J.F., Smarr, L.L., \&  Wilson, J.R. 1985, ApJS, 55, 211\\
Hua, X., \& Titarchuk, L.G. 1995, ApJ, 449, 188 (HT95)\\
Illarionov, A.F., Kallman, T., McCray, R., \& Ross, R.R. 1979, ApJ. 228, 279\\
Inoue, H. 1992, in {\it Frontiers of X-ray Astronomy}, ed. Y. Tanaka \& 
K. Koyama, (Tokyo: University Press), 291 (I92)\\
Kallman, T.R., \& Krolik, J. H. 1986, ApJ., 308, 805\\
Kallman, T. R., \& McCray, H. 1982, ApJ Supp. Ser, 50, 263\\
Liang, E.P., \& Thompson, K.A. 1980, ApJ, 240, 271\\
Lang, K.R., 1980, Astrophysical Formula, (Springer Verlag, New York)\\
Lybarskii, Yu.E., \& Sunyaev, R.A. 1982, Soviet Astr. Lett. 8, 330\\
Magdziarz, P. \& Zdziarski, A.A., 1995, MNRAS, 273, 837\\
Malkan, M. 1982, ApJ, 254, 22\\
Mauche, C.W., Raymond, J.C., \& Mattei, J.A. 1995, ApJ (in press)\\
Molteni, D., Sponholz, H., \&  Chakrabarti, S.K. 1995, ApJ (submitted)\\
Molteni, D., Lanzafame, G., \&  Chakrabarti, S.K. 1994, ApJ, 425, 161 (MLC94)\\
Muchotrzeb, B., \& Paczy\'nski, B. 1982, Acta Astron. 32, 1\\
Miyamoto, S. et al. 1991, ApJ, 383, 784
Narayan, R., \&  Yi, I. 1994, ApJ, 428, L13\\
Novikov, I., \& K.S. Thorne. 1973. in: Black Holes,
eds. C. DeWitt and B. DeWitt (Gordon and Breach, New York)\\
Paczy\'nski B., \& Bisnovatyi-Kogan, G. 1981, Acta Astron. 31, 283\\
Paczy\'nski B., \& Wiita, P.J. 1980, A\&A, 88, 23\\
Parmar, A.N. et al. 1993, A\&A, 279, 179
Perola, G.C. et al. 1986 ApJ, 306, 508\\
Peterson, B.M. et al. 1991 ApJ, 368, 119\\
Pietrini, P., \& Krolik, J.H.  1995, ApJ, in press\\
Piro, L., Yamauchi, M., \& Matsuoka, M. 1990, ApJ, 360, L35\\
Rajeev, M.R., Chitnis, V.R., Rao, A.R. \& Singh, K.P. 1994, ApJ., 424, 376\\
Rees, M.J., Begelman, M.C., Blandford, R.D. \& Phinney, E.S. 
1982, Nat, 295, 17\\
Riffert, H. 1988. ApJ, 327, 760\\
Ryu, D., Brown, G., Ostriker, J., \& Loeb, A. 1995, ApJ, (in press)\\
Shakura, N.I., \& Sunyaev, R.A. 1973, A\&A, 24, 337\\
Shakura, N.I., \& Sunyaev, R.A. 1976, MNRAS, 175, 613\\
Shapiro, S.L., \& Teukolsky, S.A., Black Holes, White Dwarfs and Neutron
Stars, 1984 (John Wiley \& Sons: New York)\\
Sobolev, V.V., 1975, Light Scattering in Planetery Atmosphere (Pargamon
Press: Oxford)\\
Sunyaev, R.A. et al., 1991, Astron. Lett., 17, 123\\
Sunyaev, R.A. et al., 1994, Astron. Lett., 20, 777\\
Sunyaev, R.A. \& Titarchuk, L.G. 1980, A\&A, 86, 121 (ST80)\\
Sunyaev, R.A. \& Titarchuk, L.G. 1985, A\&A, 143, 374 (ST85)\\
Sun, W.H. \&  Malkan, M. 1989, ApJ, 346, 68\\
Shapiro, S.L., Lightman, A.P., \& Eardley, D.M., 1974, ApJ, 204, 187\\
Tanaka, Y., 1989, in the {\it Proceedings of the 23rd ESLAB symposium}, 
ed. J. Hunt \& B. Battrick vol. 1, p.3 (Paris:ESA) (TAN) \\
Tanaka, Y. 1994, in {\it New Horizons of X-ray Astronomy},
eds. F. Makino \& T. Ohashi (Tokyo: Universal Academy Press)\\
Tanaka, Y. et al. 1995, Nat., 375, 659\\
Titarchuk, L.G. 1987, Soviet Astrofizika, 26, 57
(Astrophysics 1988, 26, 97) (T87) \\
Titarchuk, L.G. 1988, Soviet Astrofizika, 29, 634
(Astrophysics 1988, 26, 97) (T88) \\
Titarchuk, L.G. 1994, ApJ, 434, 570 (T94)\\
Titarchuk, L.G., \&  Lyubarskij, Yu 1995, ApJ, in press (TL95)\\
Titarchuk, L.G., Mastichiadis, A. \& Kylafis, N., 1996,
ApJ, submitted (TMK96)\\
Ueda, Y., Ebisawa, K., \& Done, C., 1994, PASJ, 46, 107\\
Wandel, A., \& Liang, E.P., 1991, ApJ, 380, 84\\
Wilson, C.K. \& Rothschild, R.F. 1983, ApJ, 274, 717
Yaqoob, T., 1992, MNRAS, 258, 198\\
Yaqoob, T. et al., 1993, MNRAS, 262, 435\\
Zdziarski, A.A. et al. ApJ, 363, L1 1990\\

\newpage

\centerline{\large FIGURE CAPTIONS}

\noindent Fig. 1: Schematic diagram of the accretion processes
around a black hole. An optically thick,
Keplerian disk which produces the soft component
is surrounded by an optically thin
sub-Keplerian halo which terminates in a standing shock
close to the black hole. The postshock flow Comptonizes soft photons 
from the Keplerian disk and radiates them as the hard component. Iron line
features may originate in the rotating winds.

\noindent Fig. 2: Proton and electron temperatures ($T_p$ and $T_e$ in $^o$K) 
inside the disk and the postshock region when ${\dot m}_d\!=\!0.001$ (solid
line ), 
$0.01$ (long-dashed line), $0.1$ (short-dashed line) and $1$ 
(dotted line). Other parameters are ${\dot m}_h\!=\!1$ and $M\!=\!5M_\odot$.
At higher accretion rates, the postshock region cools faster and the
convergent flow regime begins.

\noindent Fig. 3(a-b): Variation of the (a) energy spectral index $\alpha$
(observed slope in the $2-50$keV region) and (b) the mean electron 
temperature (in keV) $T_e$ as functions of the disk and halo accretion rates.
Notice the stability of $\alpha$  in the hard state and the transition 
of states when ${\dot m}_d \sim 0.1-1.0$. $M\!=\!5M_\odot$ is chosen.
Halo rates are marked on the curve. In (a), computed spectral index in
soft states due to convergent flow is also provided. Dashed curves
indicate regions where both components could contribute. In (b),
short-dashed, solid and long-dashed curves are for halo rates
$0.5$, $1.0$ and $2.0$ respectively.

\noindent Fig. 4(a-b): Same as Figs. 3a and 3b except that 
$M\!=\!10^7M_\odot$ is chosen. In this case, the average electron 
temperature is generally lower, and the spectral index generally higher.

\noindent Fig. 5: Contributions of various components to the net 
spectral shape (solid). Long dashed, short dashed, dotted and dash-dotted
curves are the contributions from the Shakura-Sunyaev disk $r>r_s$,
the reprocessed hard radiation by the Shakura-Sunyaev disk, reprocessed
soft-radiation by the postshock disk $r<r_s$ and the hard
radiation reflected from the Shakura-Sunyaev disk along the observer
($\mu\!=\!0.4$). The contribution of the preshock halo which is
orders of magnitude lower is ignored. Parameters are ${\dot m}_d\!=\!0.1$,
${\dot m}_h\!=\!1$, and $M\!=\!5\ M_\odot$.

\noindent Fig. 6: Variation of the spectral shape as the accretion rate
of the disk is varied. ${\dot m}_d\!=\!0.001$ (solid line), 
$0.01$ (long-dashed line),
$0.1$ (short-dashed line), and $1$ (dotted line). 
Though the soft luminosity increases with ${\dot m}_d$,
the hard luminosity may show `pivoting' or crossing as intermediate
energies. Note the correlation and anticorrelation of the hard flux and the
spectral index before and after the pivoting point. The dash-dotted
curve represents the hard component from convergent inflow near the 
black hole and has the characteristics of the slope $\sim 1.5$
in soft state.

\noindent Fig. 7: Spectral shape of a Keplerian disk with a 
very high disk accretion rate. The postshock region cools immediately due
to Comptonization and the cooler photons are subsequently Comptonized
by momentum deposition from the converging bulk 
motion close to the black hole. The spectral slope is $\alpha \sim 1.5$,
an universal feature of black hole candidates in the soft state.

\noindent Fig. 8(a-b): Variation of the (a) temperature (in $^o$K)
and (b) spectral shape as the accretion rate of the halo is
changed. ${\dot m}_h\!=\!1$ (solid), ${\dot m}_h\!=\!2$ (long dashed) and 
${\dot m}_h\!=\!0.5$ (short dashed). The hard flux goes up but the spectrum
hardens with ${\dot m}_h$ while the 
disk temperature and the soft luminosity remain virtually unchanged.

\noindent Fig. 9: Fate of a power-law hard component in presence
of a wind of $\tau \sim 3$ with a hundred percent convering
factor. Emitted (solid line), down-scattered component (long-dashed line)
and its absorption by the cooler outer region (short-dashed line) are 
shown.  This ``transmission bump" from the wind is indistinguishable 
from the ``reflection bump" from the disk, except for the higher 
equivalent width in the former case.

\noindent Fig. 10: Ratio of the disk angular momentum to Keplerian
angular momentum in three different cases. In case marked `A', the flow becomes
sub-Keplerian before becoming super-Keplerian close to the black hole.
In case marked `B', the disk becomes sub-Keplerian and remained so
before entering the black hole. In case marked `C', the disk
becomes Keplerian, then passed through a standing shock  at $r=13.9r_g$
before entering the black hole. 

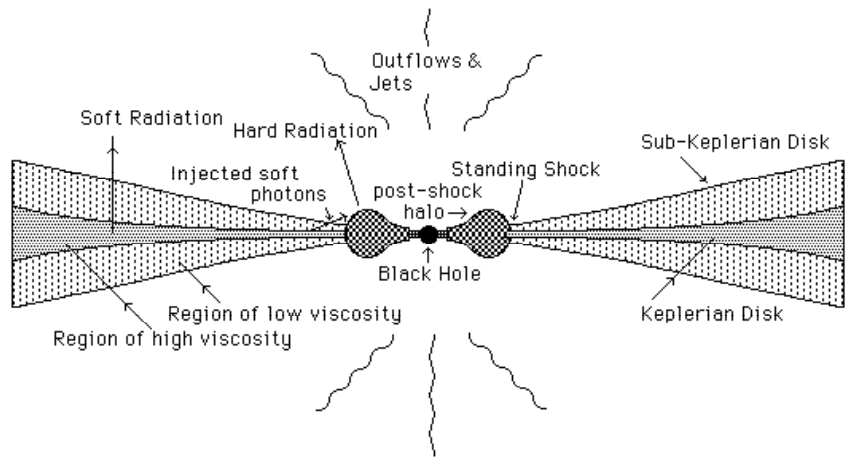
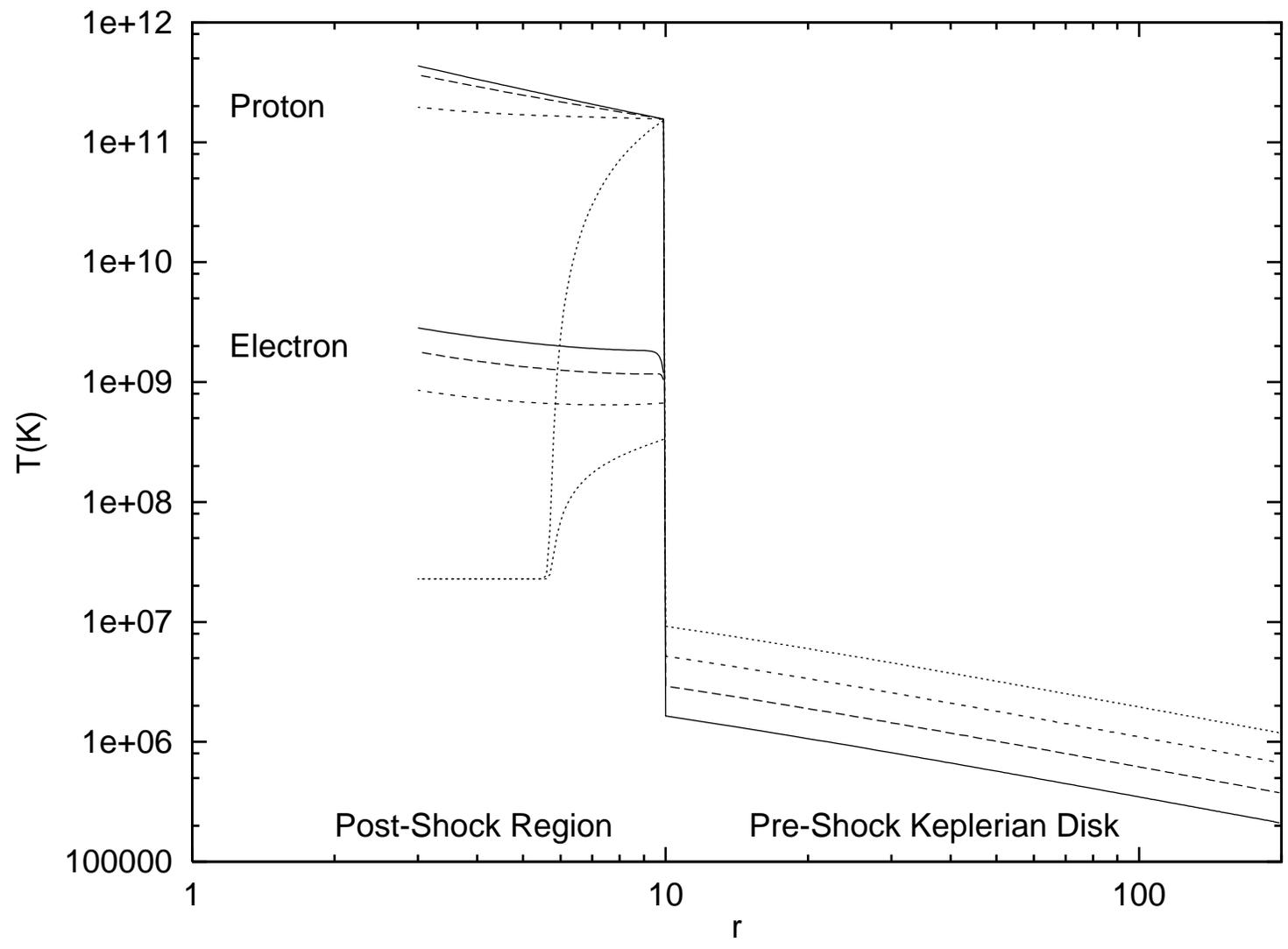
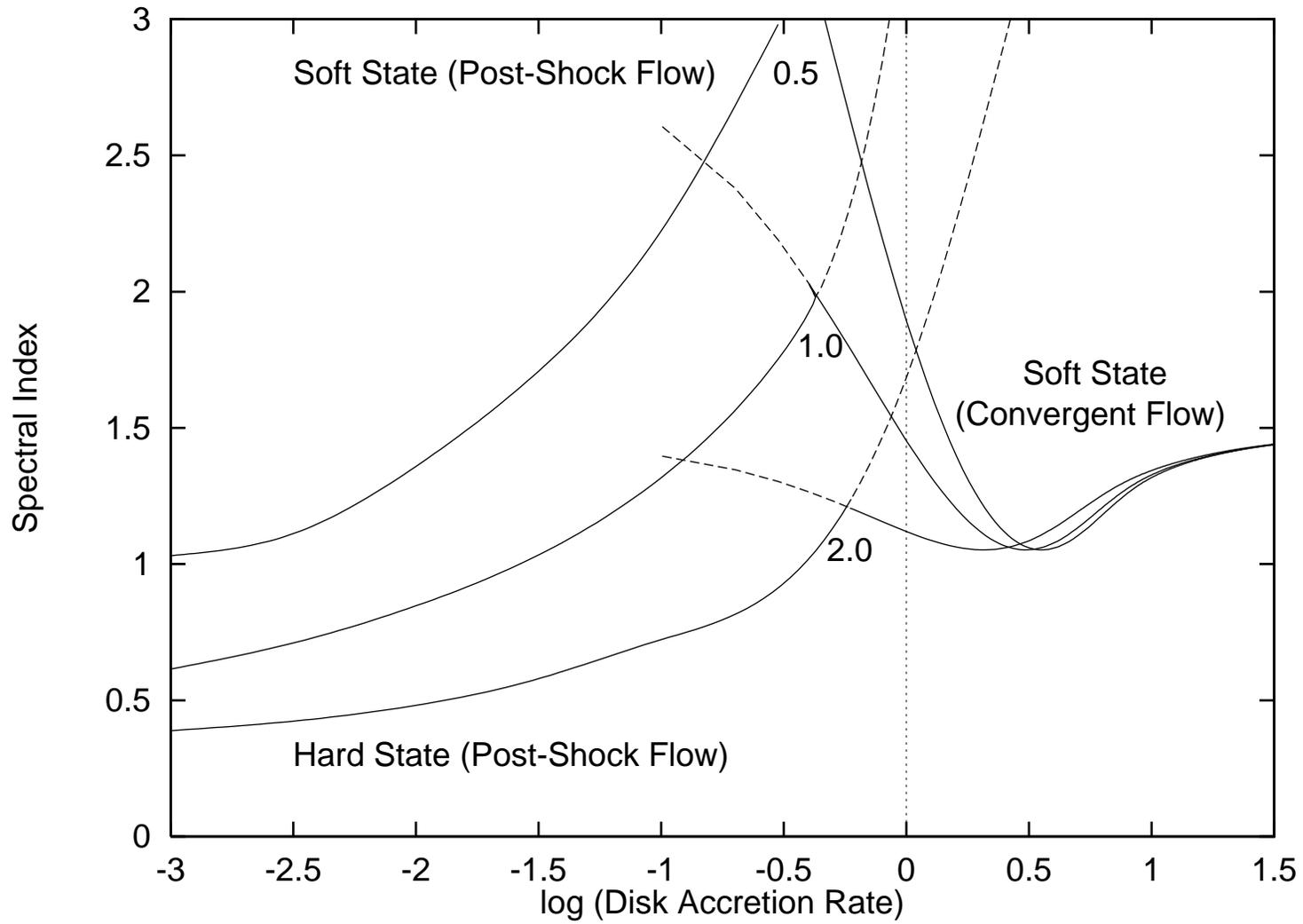
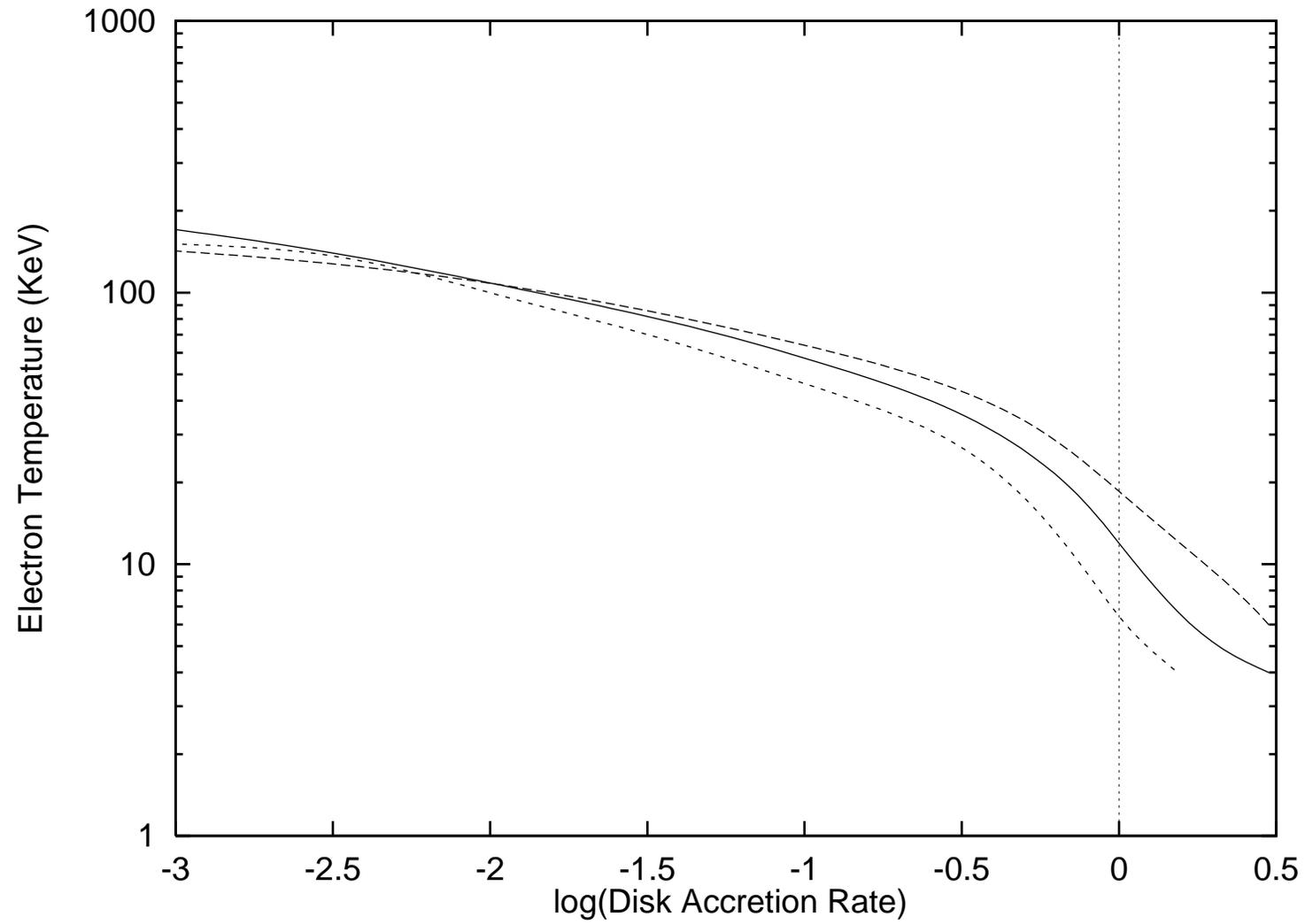
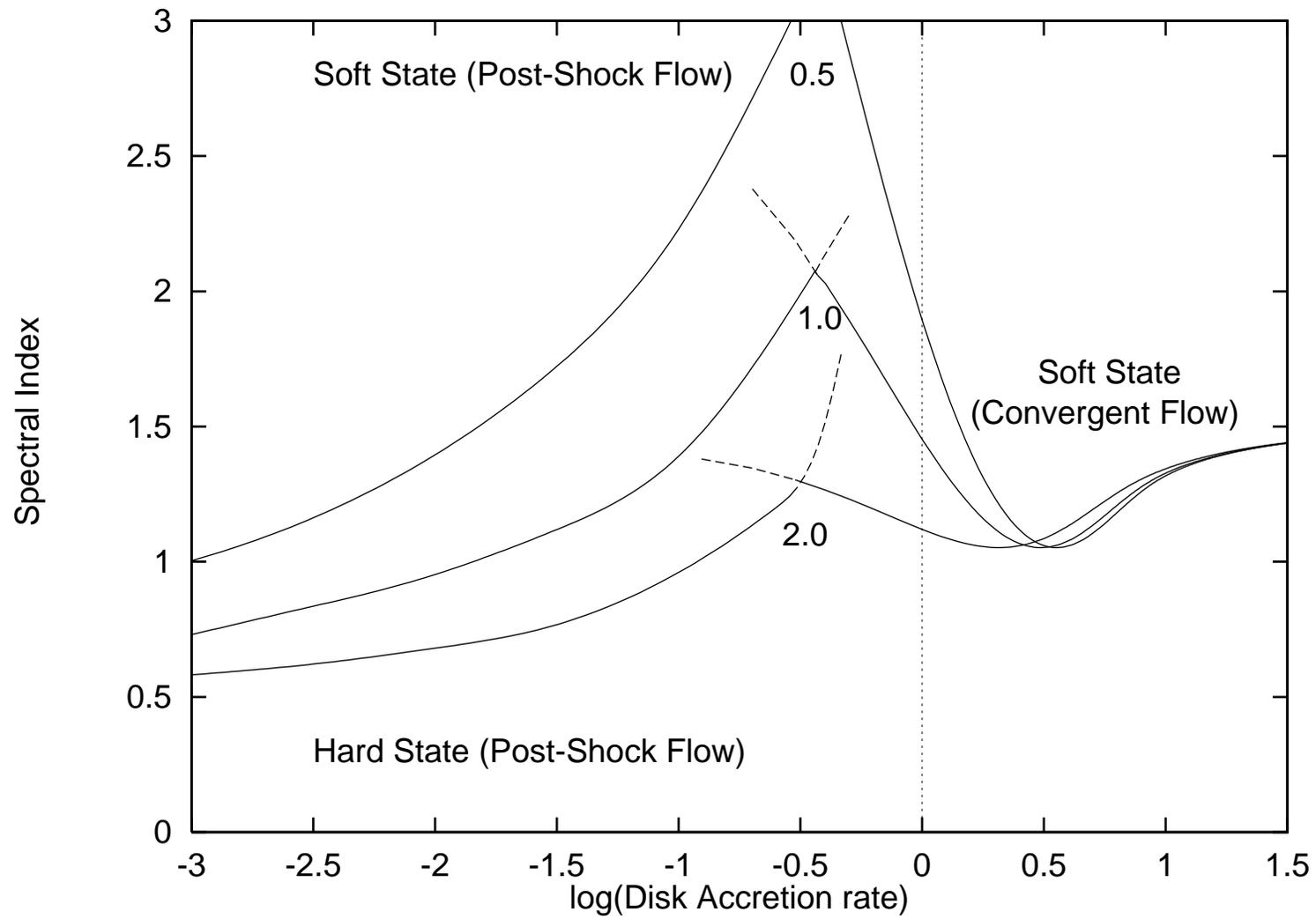
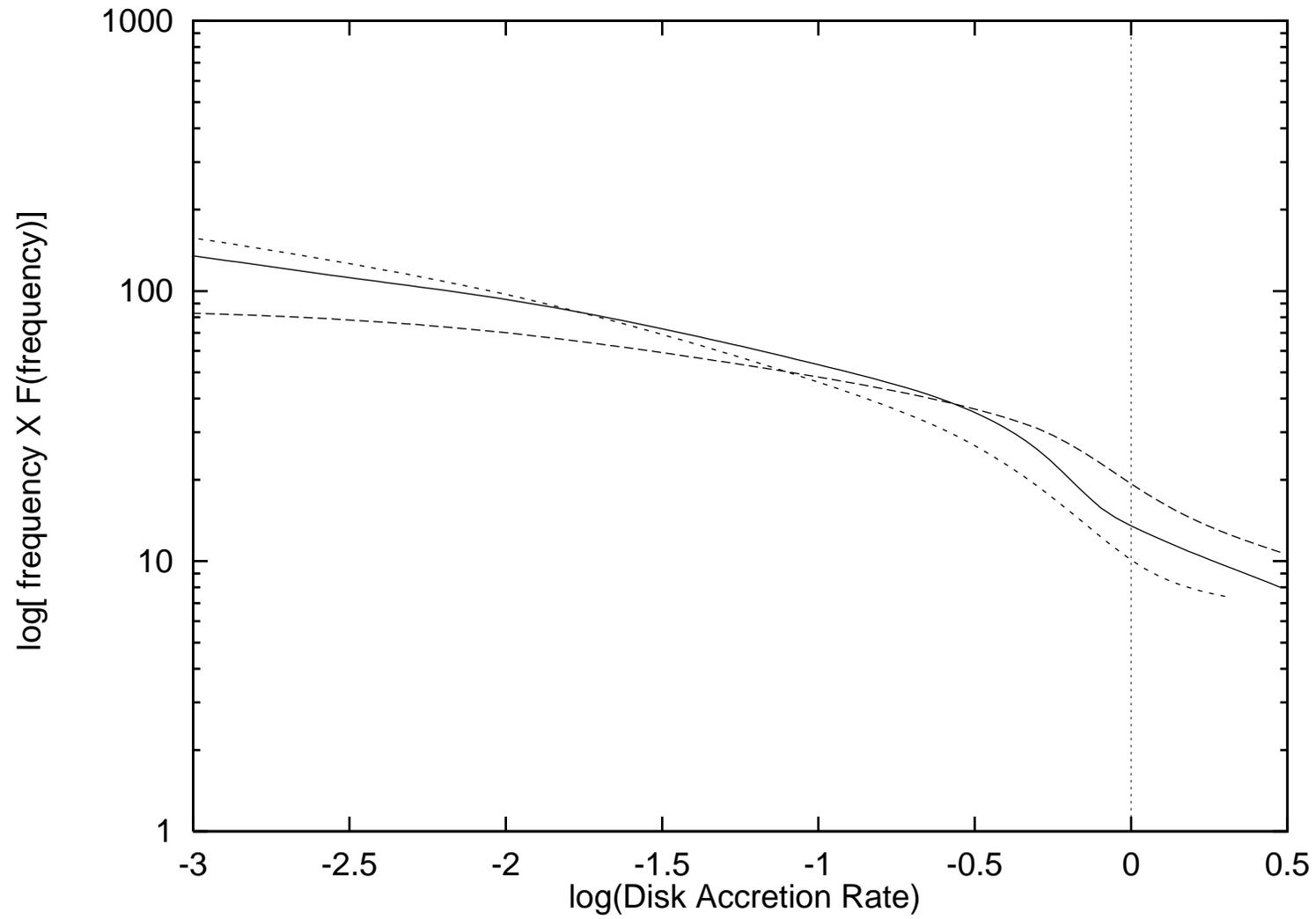
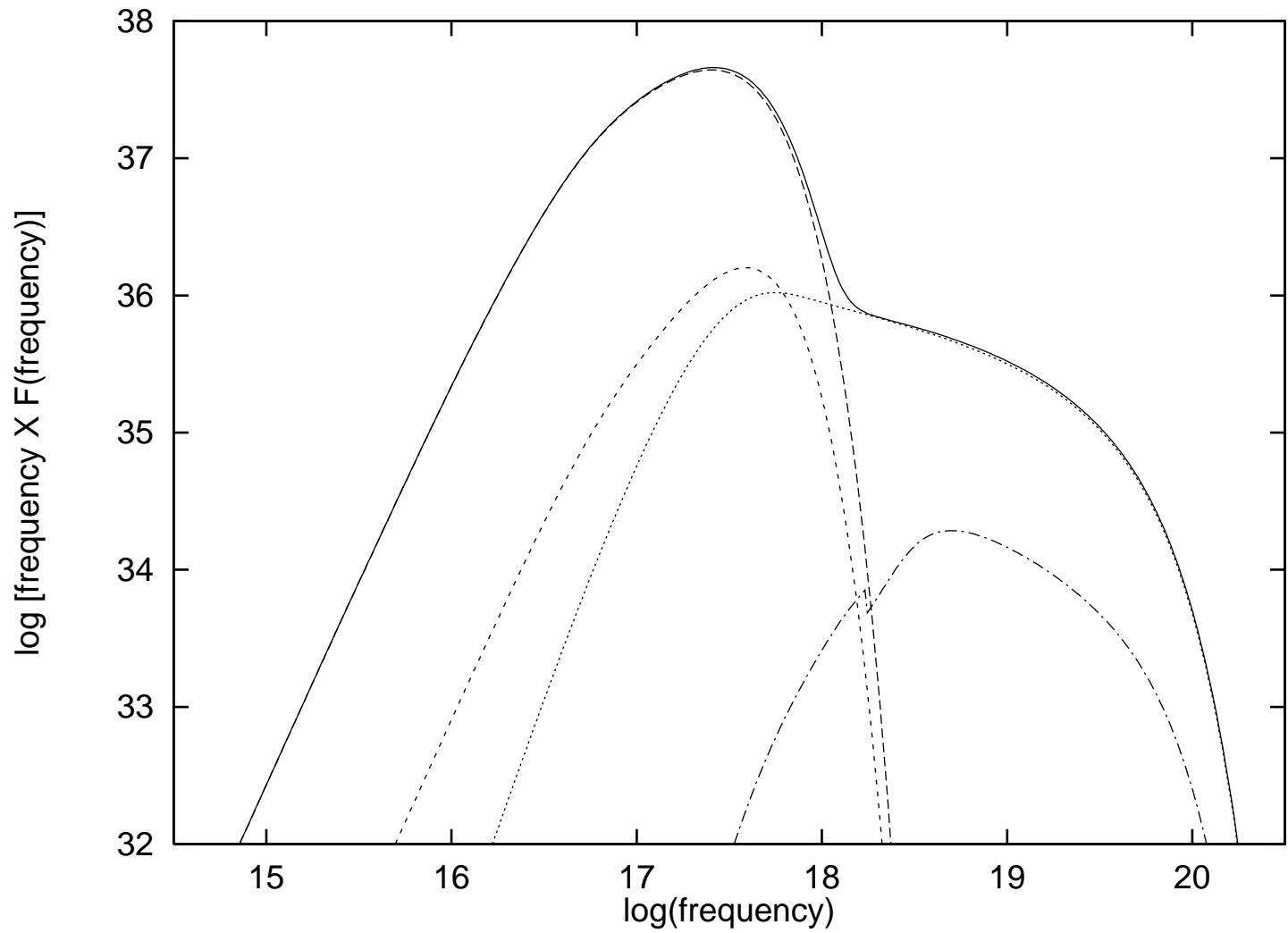
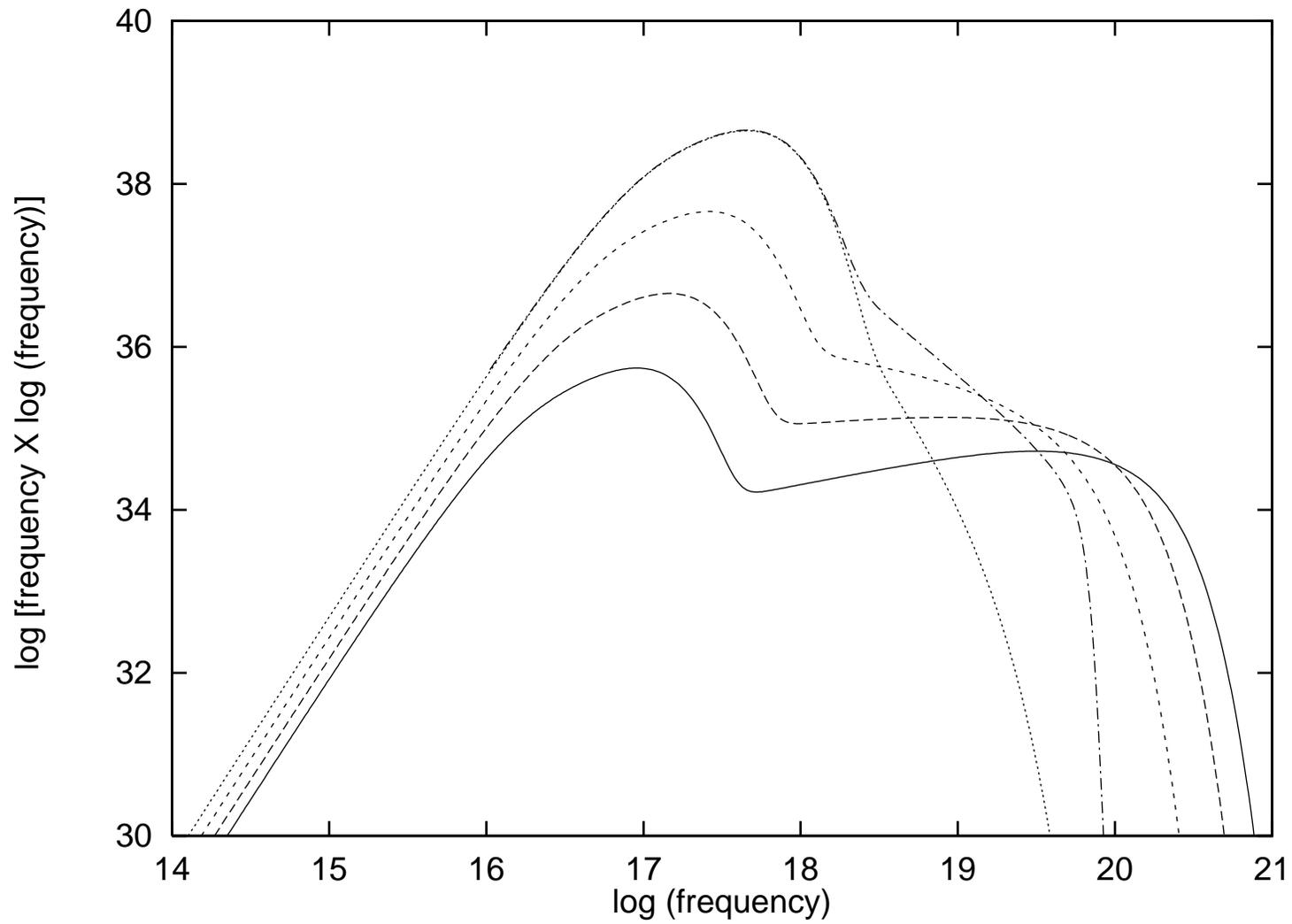
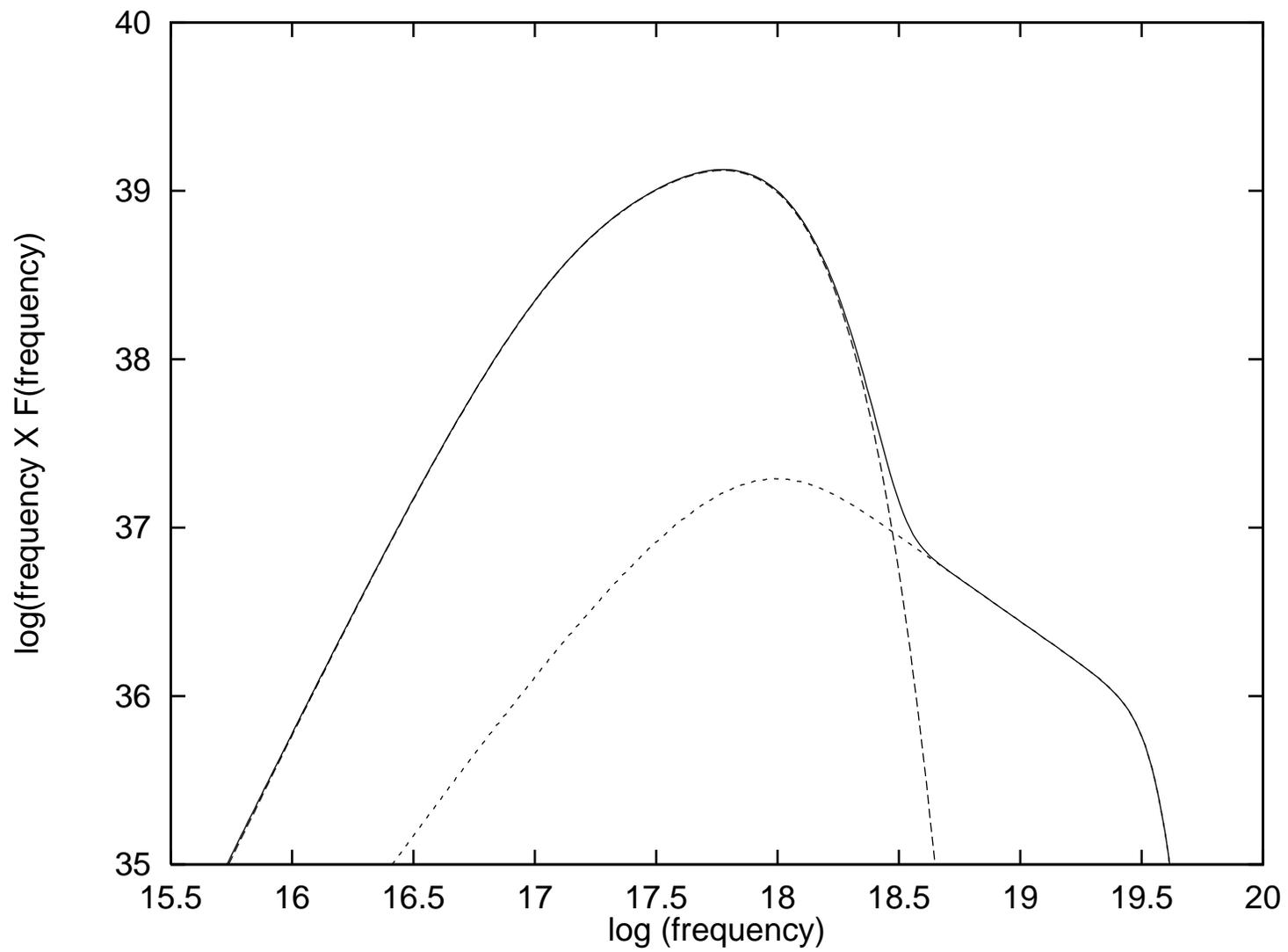
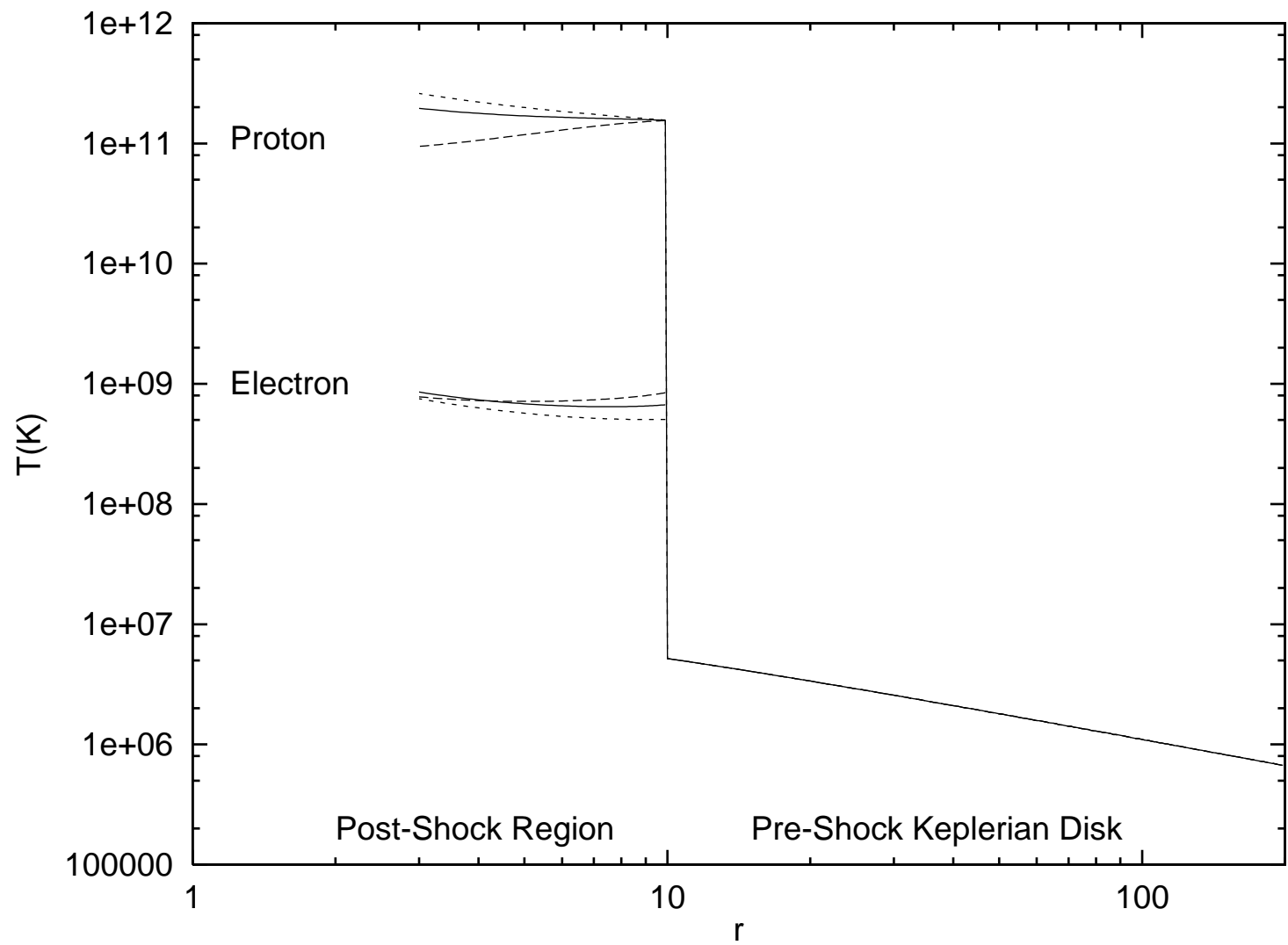
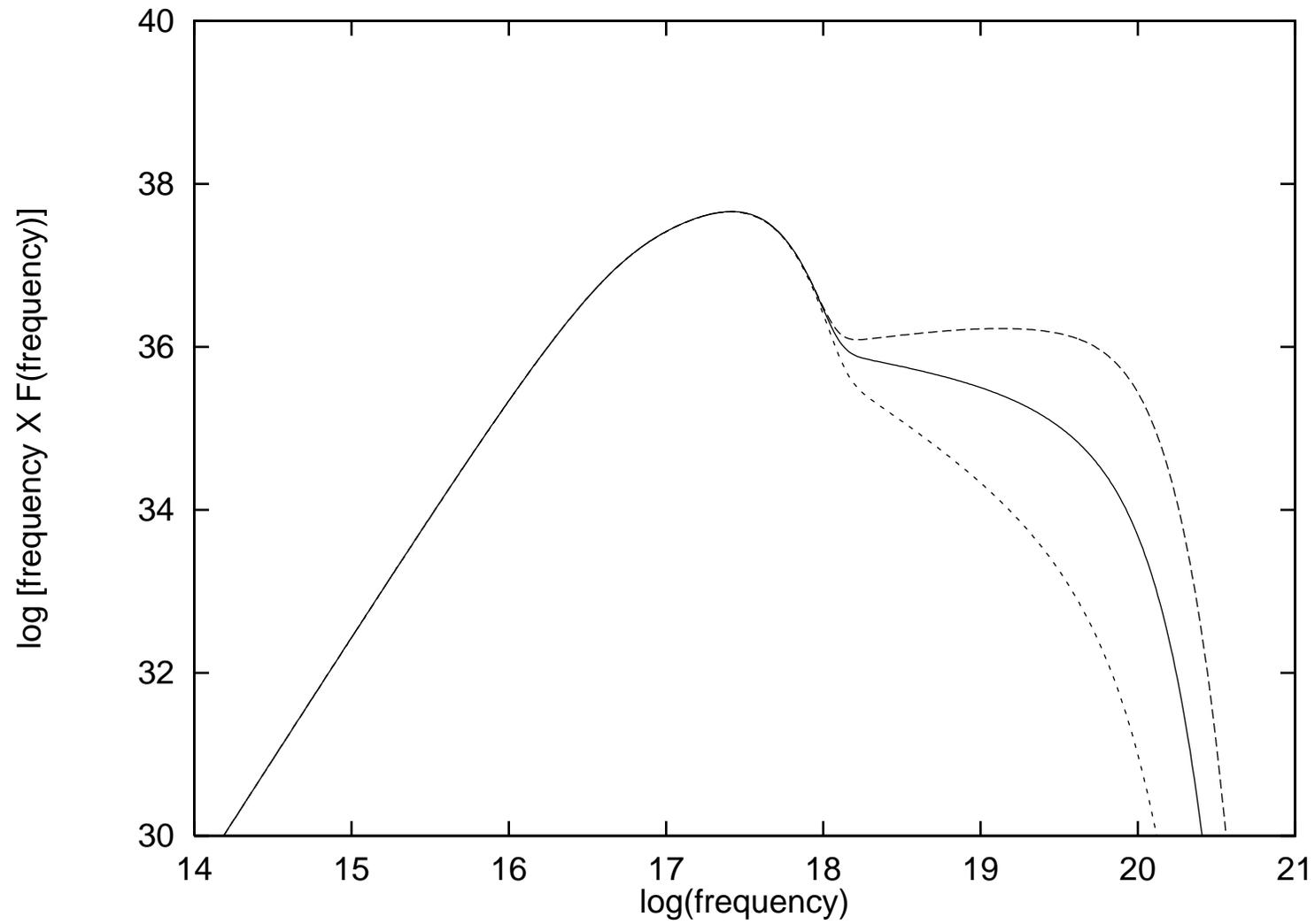
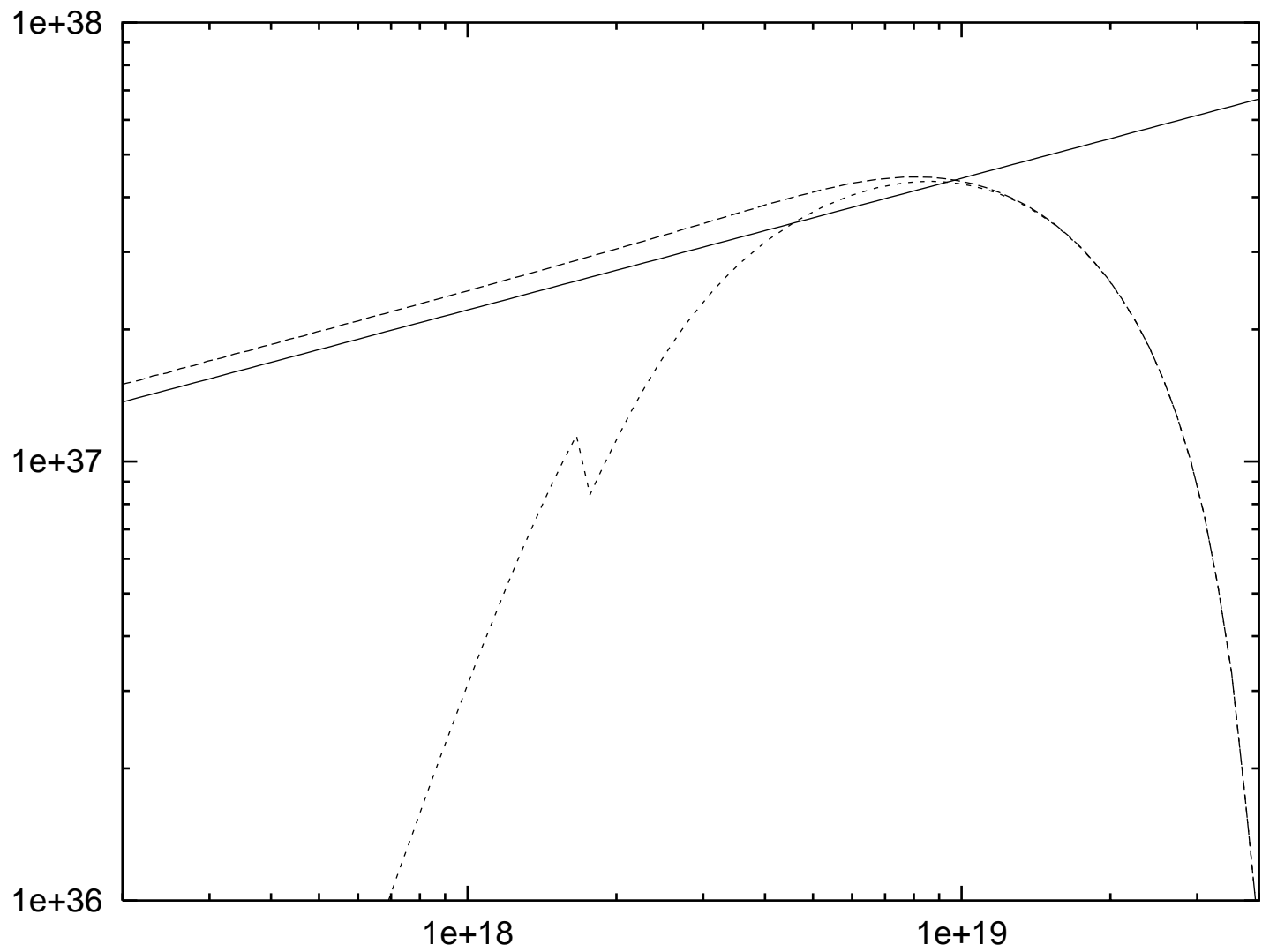
\end{document}